\documentclass[aps,pra,twocolumn,showpacs,superscriptaddress,shortbibliography]{revtex4-2}
\usepackage{graphicx} %Include figure files,superscriptaddress
\usepackage{amsmath}
\usepackage{soul}
\usepackage{graphicx,epstopdf}
\usepackage{gensymb}
\usepackage[dvipsnames]{xcolor}
\usepackage{footnote}
\usepackage{multibib}

\newcommand{\be}{\begin{equation}}
	\newcommand{\ee}{\end{equation}}
\newcommand{\bea}{\begin{eqnarray}}
	\newcommand{\eea}{\end{eqnarray}}
\newcommand{\bse}{\begin{subequations}}
	\newcommand{\ese}{\end{subequations}}

\graphicspath{{../}}
\usepackage{color}
\usepackage[colorlinks,bookmarks=false,citecolor=darkblue,linkcolor=red,urlcolor=blue]{hyperref}

\definecolor{darkred}{rgb}{0.7,0.0,0.0}

\definecolor{darkblue}{rgb}{0,0.02,0.45}

\definecolor{darkgreen}{rgb}{0.02,0.45,0.0}

\definecolor{violet}{rgb}{0.8,0.2,0.6}

\begin{document}
\title{Magnetic and crystal electric field excitations in a spin-orbit coupled frustrated hyperkagome magnet Nd$_3$Li$_3$W$_2$O$_{12}$}
\author{R. Kolay}
\thanks{These authors contributed equally to this work.}
\affiliation{School of Physics, Indian Institute of Science Education and Research Thiruvananthapuram-695551, India}
\author{S. Guchhait}
\thanks{These authors contributed equally to this work.}
\affiliation{Department of Physics, Technical University of Denmark, 2800 Kongens Lyngby, Denmark.}
\author{Indrajeet S. Rathore}
\affiliation{School of Physics, Indian Institute of Science Education and Research Thiruvananthapuram-695551, India}
\affiliation{Jülich Centre for Neutron Science for Quantum Materials and Collective Phenomena (JCNS-2), Forschungszentrum Jülich GmbH, 52425 Jülich, Germany}
\affiliation{RWTH Aachen, Faculty of Mathematics, Computer Science and Natural Sciences, 52074 Aachen, Germany}
\author{S. Nandi}
\affiliation{Jülich Centre for Neutron Science for Quantum Materials and Collective Phenomena (JCNS-2), Forschungszentrum Jülich GmbH, 52425 Jülich, Germany}
\affiliation{RWTH Aachen, Faculty of Mathematics, Computer Science and Natural Sciences, 52074 Aachen, Germany}
\author{M. D. Le}
\affiliation{ISIS Neutron and Muon Source, Science and Technology Facilities Council, Rutherford Appleton Laboratory, Didcot OX11 0QX, United Kingdom}
\author{N. B. Christensen}
\email{nbch@fysik.dtu.dk}
\affiliation{Department of Physics, Technical University of Denmark, 2800 Kongens Lyngby, Denmark.}
\author{R. Nath}
\email{rameshchandra.nath@gmail.com}
\affiliation{School of Physics, Indian Institute of Science Education and Research Thiruvananthapuram-695551, India}
\date{\today}
	
\begin{abstract}
Rare-earth based garnets provide a viable platform for studying the frustrated driven magnetic properties of the hyperkagome lattices. Herein, we report a comprehensive study of the magnetic properties and crystal electric field (CEF) scheme of a new Nd$^{3+}$ based hyperkagome antiferromagnet, Nd$_3$Li$_3$W$_2$O$_{12}$ belonging to the garnet family via magnetization, heat capacity, and inelastic neutron scattering (INS) measurements. Magnetization measurement reveals a dominant antiferromagnetic interaction with a low temperature Curie-Weiss temperature $\theta_{\rm CW}^{\rm LT} \simeq -0.2$~K. Two broad maxima are observed in the magnetic heat capacity data under magnetic fields, implying multilevel Schottky anomalies due to the effect of CEF and display a two-step magnetic entropy release. No magnetic long-range order is observed down to 0.1~K. The CEF excitations of the Nd$^{3+}$ ($J=9/2$) ion with $D_2$ point group symmetry, probed via INS experiments, show non-dispersive excitations characterizing the transitions among the CEF energy levels. The simultaneous fit of the INS spectra at different temperatures enabled the mapping of the CEF Hamiltonian and the energy eigenvalues of the Kramers' doublets. The simulation using the obtained CEF parameters reproduces the experimental magnetic susceptibility, magnetic isotherms, and magnetic heat capacity data. The thermodynamic properties and INS-derived crystal-field scheme confirm a Kramers' doublet ground state with an effective spin $J_{\rm eff} = 1/2$ at low temperatures.
\end{abstract}

\maketitle

\section{Introduction}
Frustrated magnetism is a fascinating area of contemporary condensed matter physics, offering a fertile platform for the exploration of emergent low temperature phenomena such as quantum spin liquids (QSLs), spin ice, and other exotic magnetic phases.~\cite{Savary016502,Bramwell1495,Starykh052502}. In particular, the frustrated magnets with rare-earth ions ($4f$) are almost an unexplored territory in which a delicate interplay among magnetic correlations, spin-orbit coupling (SOC), and crystal electric field (CEF) effects gives rise to a rich variety of quantum phenomena~\cite{Rau357,Graham166703,Bordelon1058,Li107202,Sibille711}. In these systems, the CEF is typically weaker than the SOC and primarily influences the low-energy eigen states. For Kramers-active ions, for example, the CEF splits the spin-orbit multiplet into a series of doublets, with the lowest-lying doublet serving as the ground state at low temperatures and behaving as an effective spin-$1/2$ ($J_{\rm eff}=1/2$)~\cite{Guchhait144434,Zhang256503,Ranjith180401,Yamamoto075114}. Furthermore, the CEF can induce strong anisotropy in the magnetic interactions and enhance quantum tunneling effects, making a detailed understanding of the CEF scheme essential for elucidating the low temperature magnetic properties and emergent quantum states in rare-earth-based frustrated magnets~\cite{Rau144417,Tomasello155120,Gao024424,Scheie144432}.

%\textbf{Over the years, numerous two-dimensional (2D) rare-earth-based frustrated magnets have been extensively studied since they harbour fascinating low temperature quantum phases as a consequence of quantum fluctuations due to reduced dimensionality and low spin value as well as magnetic frustration. A few examples include gapless QSL in YbMgGaO$_4$~\cite{Li16419}, a Dirac spin liquid in YbZn$_2$GaO$_5$~\cite{Bag266703}, spin-liquid in triangular-honeycomb antiferromagnet TbInO$_3$~\cite{Clark262}, unconventional spin fluctuation in YbBO$_3$~\cite{Somesh064421}, spinon excitation in CeMgAl$_{11}$O$_{19}$~\cite{Gaoeaed7778}, multipolar excitations in NaErSe$_2$~\cite{Zhang256503}, unconventional low temperature spin dynamics in BiErGeO$_5$~\cite{Mohanty2026}, and an emergent order in Ising magnet Dy$_3$Mg$_2$Sb$_3$O$_{14}$~\cite{Paddison13842}.}

As compared to the two-dimensional (2D) frustrated magnets~\cite{Bag266703,Clark262,Somesh064421,Mohanty214452}, rare-earth-based magnets with three-dimensional (3D) frustrated lattices have received considerably less attention. This is primarily due to two factors: (i) quantum fluctuations are generally suppressed in higher dimensions, making 3D magnets appear less favorable for hosting exotic quantum states, and (ii) identifying suitable materials that exhibit strong magnetic frustration in 3D is inherently challenging. In this regard, a few compounds with frustrated 3D systems are thoroughly exploited. For instance, the pyrochlore oxides with the general formula $R_2B_2$O$_7$ ($R$ is a trivalent rare earth ion and $B$ a tetravalent transition metal ion)~\cite{Gao1052,Kermarrec14810,Hallas105,Gardner53} and the garnets with the general formula $R_3M_2X_3$O$_{12}$ ($M$ = Ga, Sc, In, Te, and $X$ = Ga, Al, Li)~\cite{Petrenko4570,Joseph179,Raymond236701,Xin014436,Cai184415,Petit013030} featuring hyperkagome geometry are reported to harbor a wide range of exciting phases encompassing spin-liquid, spin-ice, etc.

Similar to other rare-earth ions, Nd$^{3+}$-based frustrated magnets have also attracted considerable attention in recent years.
%Being a Kramers' active ion, Nd$^{3+}$ often realizes $J_{\rm eff} = 1/2$ ground state at lower temperatures which along with geometric frustration leads to enhanced quantum fluctuations and hence, intriguing magnetic ground states. However, only a handful of experimental Nd$^{3+}$-based frustrated systems have been identified so far.
Notable examples include QSL in Ising antiferromagnet NdTa$_7$O$_{19}$~\cite{Arh2022}, a long-range all-in/all-out antiferromagnetic order below 0.4~K in Nd$_2$Zr$_2$O$_7$ revealed via neutron diffraction~\cite{Xu224430}, an Ising-like ground state observed in Nd$_3$Ga$_5$O$_{12}$~\cite{Zhao014441}, persistent spin fluctuations in NdZnAl$_{11}$O$_{19}$~\cite{Cao144409}, coexistence of short-range magnetic correlations with magnetic long-range order (LRO) at $T_{\rm N} \simeq 0.26$~K in Nd$_2$Be$_2$GeO$_7$~\cite{Liu184413}, and the absence of magnetic LRO in the square lattice antiferromagnet NdKNaNbO$_5$~\cite{Guchhait144434}. Therefore, the search for new Nd$^{3+}$-based frustrated magnets remains highly desirable, as the interplay between magnetic frustration and SOC-induced anisotropic interactions may stabilize novel quantum states.

\begin{figure}
\includegraphics[width=\columnwidth] {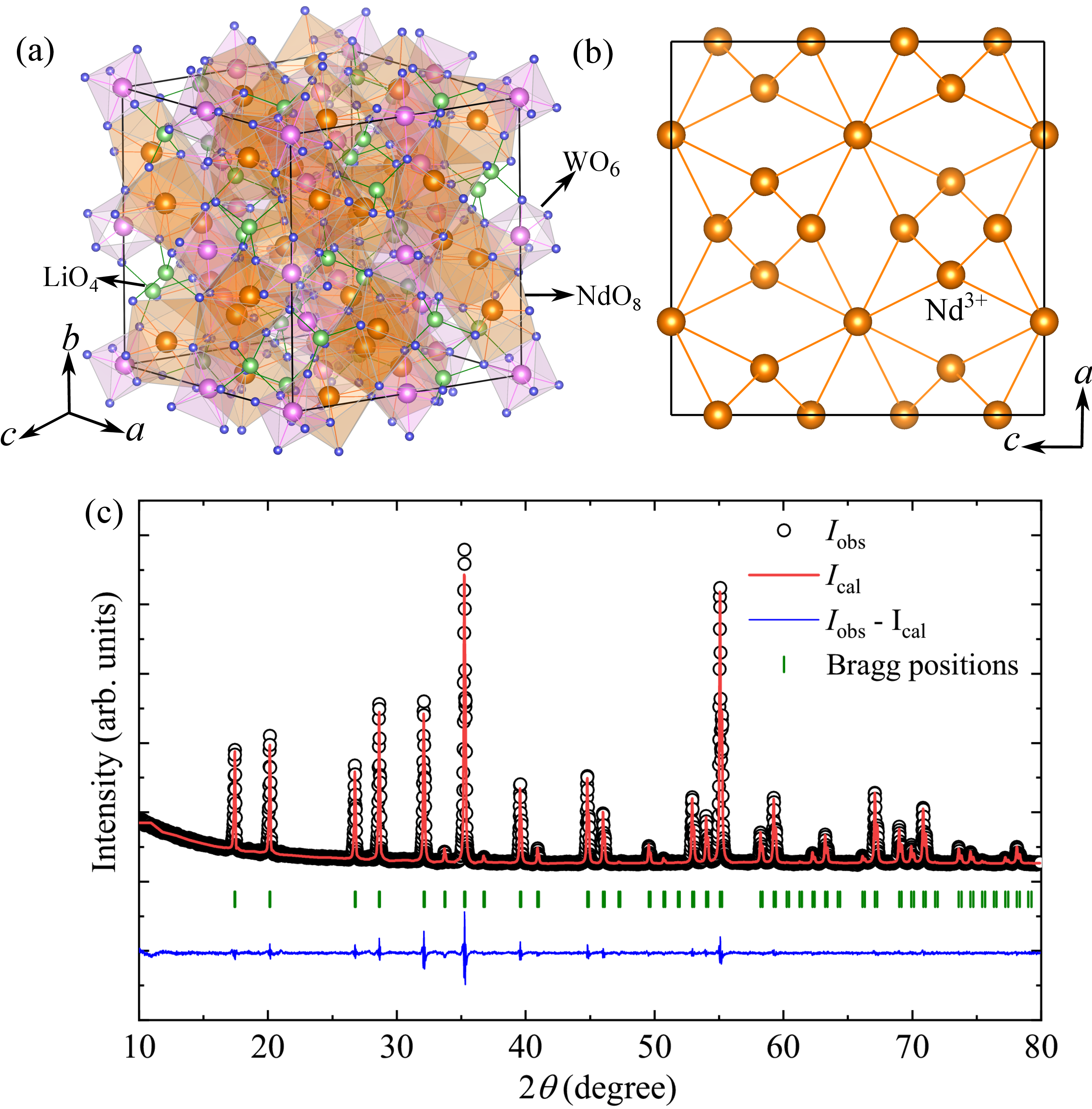}
\caption{\label{Fig1}(a) A 3D view of the crystal structure of Nd$_3$Li$_3$W$_2$O$_{12}$ formed by corner/edge sharing NdO$_8$ polyhedral, WO$_6$ octahedral, and LiO$_4$ tetrahedral units.
(b) The hyperkagome lattice formed by the Nd$^{3+}$ ions projected in the $ac$-plane. (c) Room temperature powder XRD pattern of Nd$_3$Li$_3$W$_2$O$_{12}$. The open black circles represent the observed intensity, and the solid red line is the Rietveld fit with $\chi^2 \simeq 4.6$. Expected Bragg positions are shown in green vertical bars, and the bottom line indicates the difference between observed and experimental intensities.}
 \end{figure}

In this paper, we report the bulk properties and CEF splitting of the 3D hyperkagome compound Nd$_3$Li$_3$W$_2$O$_{12}$ that crystallizes in a cubic space group $Ia\bar{3}d$. The 3D crystal structure of the titled compound is depicted in Fig.~\ref{Fig1}(a) where the NdO$_8$ polyhedra are connected via WO$_6$ octahedra and LiO$_4$ tetrahedra. To simplify the spin lattice, we have shown the connectivity among Nd$^{3+}$ ions in Fig.~\ref{Fig1}(b), removing all other atoms. The Nd$^{3+}$ ions are decorated on corner-shared triangular motifs and constitute a hyperkagome network. Except for the crystal structure, no other properties are reported for this compound~\cite{Cussen1832}. Our magnetization and heat capacity measurements suggest a Kramers' doublet with $J_{\rm eff} = 1/2$ ground state. No magnetic LRO is detected down to $100$~mK. We successfully mapped out the CEF energy levels by modeling the inelastic neutron scattering spectra. Finally, using these energy eigenvalues, we simulated the temperature and field-dependent magnetization and heat capacity, which replicate our experimental data.
%\textbf{However, the small deviation of the experimental data from the simulation implies the existence of low-energy interactions among the spins.}

%In this context, the rare-earth garnet family with the general formula $A_3B_2C_3O_{12}$ ($A$= Nd, Gd, Tb, Dy, Ho, Yb, $B$ = Ga, Sc, In, Te, $C$ = Ga, Al, Li) delivers a panoply of compound features exciting magnetic phases. The most celebrated compound in this family is Gd$_3$Ga$_5$O$_{12}$, as it hosts a spin liquid state above a freezing temperature $T_{\rm g}\sim 0.14$~K~\cite{}. However, the most recent study proposed a hidden multipolar long-range order emerging from 10 spin loops~\cite{}.  
%In thsi context, rare-earth ion with pyrochlore and hyperkagome dtructure delivers a wide rnage .....Bose Einstein condensation in Yb$_2$Si$_2$O$_7$~\cite{Hester027201},

\section{Experimental Details}
The polycrystalline sample of Nd$_3$Li$_3$W$_2$O$_{12}$ was synthesized using the standard solid-state reaction method. A stoichiometric amounts of Li$_2$CO$_3$ (Sigma Aldrich, $\geq 99.9$~\%), Nd$_2$O$_3$ (Sigma Aldrich, $\geq 99.9$~\%), and WO$_2$ (Sigma Aldrich, $\geq 99.99$~\%) were taken in a mortar pestle and ground thoroughly for several hours. To remove the moisture, Nd$_2$O$_3$ was pre-fired at $600^\circ$~C for 12~hrs. Considering the volatility of Li$_2$CO$_3$, we used an excess of 5~\% of it. The grounded powder was pressed into pellets, placed in an alumina crucible, and fired at $750\degree$~C for 48~hrs with intermittent grindings. The phase purity of the synthesized sample was confirmed by doing powder x-ray diffraction (XRD) in a PANalytical x-ray diffractometer (Cu$K_\alpha$, $\lambda_{\rm avg}= 1.5418$ \AA) at room temperature. Figure~\ref{Fig1}(c) portrays the powder XRD pattern at room temperature along with the Rietveld fit using the \texttt{FULLPROF} software~\cite{Carvajal55}. All the Bragg peaks are generated using a cubic structure with space group $Ia\bar{3}d$. We did not observe any additional peaks, that suggests the phase purity of the compound. The Rietveld fit returns the following lattice parameters $a = b = c =  12.467(3)$~\AA, and unit cell volume $V_{\rm cell} = 1937.83(1)$~\AA$^3$. These values are in good agreement with the previously reported data~\cite{Cussen1832}.

Magnetization ($M$) as a function of temperature ($T$) and magnetic field ($0\leq\mu_0H\leq 9$~T) was measured down to 1.8~K using a vibrating sample magnetometer (VSM) attachment to the Physical Property Measurement System (PPMS, Evercool-II, Quantum Design). %Isothermal magnetization [$M(H)$] data were recorded at various temperatures from 0 to 7~T. 
The temperature-dependent heat capacity [$C_{\rm p}(T)$] was measured in a wide temperature range ($0.1$~K$\leq T\leq300$~K) on a small sintered pellet using the thermal relaxation technique in PPMS by varying the magnetic field from 0 to 9~T. For attending the temperature below 2~K, a dilution insert was used in a PPMS (Dynacool, Quantum Design).

Zero-field inelastic neutron scattering (INS) measurements were performed on the direct-geometry time-of-flight spectrometer MARI~\cite{Le168646} at the ISIS Neutron and Muon Source, Rutherford Appleton Laboratory, UK. Approximately 3~g of Nd$_3$Li$_3$W$_2$O$_{12}$ powder was loaded into annular Al sample cans and cooled using a top-loading closed-cycle refrigerator. The data were collected at 6, 100, and 250~K with incident neutron energies $E_i=29.7$, 60, and 180~meV and Gd chopper frequency of 400~Hz. The elastic energy resolutions correspond to the incident neutron energies of 0.5, 1.1, and 6~meV, respectively. The data reduction and analysis were carried out using the MANTID software package~\cite{Arnold156}.

\section {Results}
%\subsection{X-ray diffraction}
%----------------------------------------------------
% \begin{figure}
% 	\includegraphics[width=\columnwidth]{Fig2}
% 	\caption{\label{Fig2} Room temperature powder XRD pattern of Li$_2$NiGe$_3$O$_8$. The open black circles represent the experimental data, and the solid red line is the Rietveld fit. Expected Bragg positions are shown in pink vertical bars, and the bottom line indicates the difference between observed and experimental intensities. 
% 	Inset: The single trillium lattice formed by Ni$^{2+}$ ions.}
% \end{figure}

\subsection{Magnetization}
\begin{figure*}
\includegraphics[width=\textwidth]{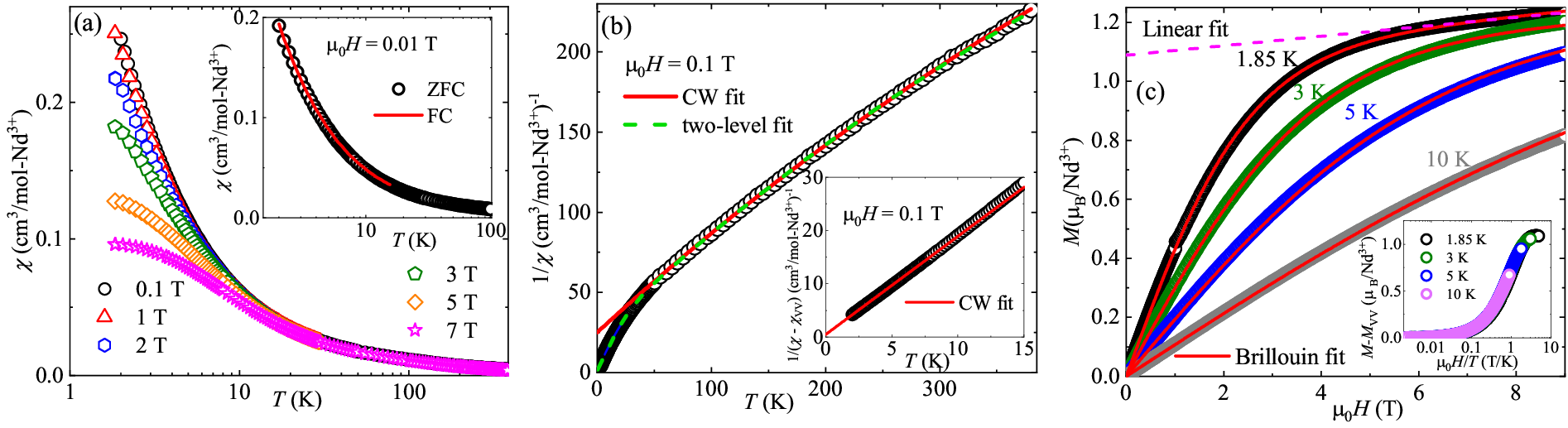}
\caption{\label{Fig2} (a) $\chi$ as a function of $T$ measured in different applied magnetic fields. Inset: $\chi(T)$ measured at $\mu_{0}H = 0.01$~T in both ZFC and FC protocols. (b) Inverse susceptibility $1/\chi(T)$ data for $\mu_{0}H = 0.1$~T. The solid and dashed lines correspond to the CW and two-level model fits, respectively. Inset: Van-Vleck corrected inverse susceptibility $1/(\chi-\chi_{\rm VV})$ vs $T$ in the low-$T$ regime. The solid line is the CW fit. (c) Magnetic isotherms measured at different temperatures. The red solid lines are the fits using Eq.~\eqref{BF}. The dashed line is the linear fit to the high field data at $T = 1.85$~K. Inset: Scaling of Van-Vleck corrected magnetization as a function of $\mu_{0}H/T$ for different temperatures.}
\end{figure*}
Temperature dependent magnetic susceptibility $\chi~(\equiv M/H)$ measured in different applied fields is depicted in Fig.~\ref{Fig2}(a). As temperature decreases, $\chi(T)$ increases systematically, without showing any magnetic LRO down to 1.85~K. However, with increasing field, $\chi$ in the low temperature regime decreases as expected and exhibits a tendency of saturation. A similar behavior is reported in several rare-earth-based magnetic materials~\cite{Guchhait144434,Guchhait214437,Sebastian104428}. $\chi(T)$ measured under zero-field-cooled (ZFC) and field-cooled (FC) protocols doesn't feature any bifurcation [see inset of Fig.~\ref{Fig2}(a)], ruling out the spin freezing or spin glass like behaviour down to 1.85~K~\cite{Kolay104403}.

The inverse susceptibility ($1/\chi$) vs $T$ in an applied field of $\mu_0H = 0.1$~T is displayed in Fig.~\ref{Fig2}(b). To extract the magnetic parameters, the high-temperature part of $1/\chi$ was fitted by the Curie Weiss (CW) law, 
\begin{equation}
	\chi(T) = \chi_{0}+\frac{C}{(T-\theta_{\rm CW})}.
\end{equation}
Here, $\chi_{0}$ is the $T$-independent susceptibility, $C$ is the Curie constant, and $\theta_{\rm CW}$ is the characteristic CW temperature. The high-$T$ fit ($T\geq 125$~K) is shown in Fig.~\ref{Fig2}(b) that returns the parameters: $\chi_{0}^{\rm HT} = 9.35(2) \times 10^{-4}$~cm$^{3}$/mol, $C^{\rm HT} = 1.45(1)$~cm$^{3}$K/mol, and $\theta_{\rm CW}^{\rm HT}= -37.21(1)$~K. From the value of $C$, the effective magnetic moment ($\mu_{\rm eff}^{\rm HT}$) is calculated using the relation $\mu_{\rm eff}^{\rm HT}= \sqrt{3k_{\rm B}C^{\rm HT}/N_{\rm A}}$ to be 3.42(1)~$\mu_{\rm B}$, where $N_{\rm A}$ is the Avogadro's number, $k_{\rm B}$ is the Boltzmann constant, and $\mu_{\rm B}$ is the Bohr magneton. This value of $\mu_{\rm eff}^{\rm HT}$ matches closely with the expected theoretical value of a free Nd$^{3+}$ ions with $J = 9/2$ and Land\`e-$g$ factor $g_J = 8/11$. Here, the large and negative value of $\theta_{\rm CW}^{\rm HT}$ does not reflect the presence of strong antiferromagnetic (AFM) interaction among the spins. It rather reflects the effect of CEF excitations at high temperatures. At high temperatures, all the higher excited Kramers' doublets are occupied and contribute to $\theta_{\rm CW}^{\rm HT}$~\cite{Guchhait144434}.

For temperatures below 50~K, $1/\chi$ changes its slope due to the depopulation of the CEF levels. Therefore, we fitted $1/\chi(T)$ by the CW law after correcting the Van-Vleck contribution ($\chi_{\rm VV}$) in the low temperature regime ($2 \leq T \leq 10$~K) to get an estimate of the intrinsic spin interaction among the Nd$^{3+}$ moments. $\chi_{\rm VV}$ was estimated for the $M$ vs $H$ analysis at $T = 1.85$~K, as discussed later. The low-$T$ CW fit is displayed in the lower inset of Fig.~\ref{Fig2}(b) which returns the parameters: $C^{\rm LT} = 0.54 (1)$~cm$^{3}$K/mol and $\theta_{\rm CW}^{\rm LT} = -0.26(1)$~K. We calculated $\mu_{\rm eff}^{\rm LT}$ using the value of $C^{\rm LT}$ which turns out to be $2.08(2)~\mu_{\rm B}$. This corresponds to an effective spin $J_{\rm eff} = 1/2$ with an average $g$-value of $2.40(1)$, which closely align with our calculated $g-$value from the INS data. A similar $g$-value is also reported for other Nd$^{3+}$ based systems~\cite{Guchhait144434}. Furthermore, the small and negative value of $\theta_{\rm CW}^{\rm LT}$ suggests that the interaction among the $J_{\rm eff} = 1/2$ Nd$^{3+}$ spins is weak but predominantly antiferromagnetic (AFM) in nature.

Since the simple CW law cannot describe the $\chi(T)$ data across the entire temperature range, we employed a two-level CW fit to capture the effects of the excited crystal-field levels. In this two-level CW model one can express $\chi (T)$ as~\cite{Guchhait214437,Sebastian104428,Mugiraneza95}:
\begin{equation}\label{two-level}
    \chi(T) = \chi_0^{\rm CEF} + \frac{1}{8(T-\theta_{\rm CW}^{\rm CEF})} \left[\frac{\mu_{\rm eff, 0}^2 + \mu_{\rm eff, 1}^2e^-(\frac{\Delta^{\rm CEF}}{k_{\rm B}T})}{1 + e^-(\frac{\Delta^{\rm CEF}}{k_{\rm B}T})}\right].
\end{equation}
Here, $\Delta^{\rm CEF}$ is the energy splitting between the ground state and the first excited Kramers' doublets, $\mu_{\rm eff, 0}$ and $\mu_{\rm eff, 1}$ represent the effective magnetic moments of the ground state and first excited state, respectively. Applying this model, we were able to fit the $1/\chi(T)$ data over the entire temperature range as shown in Fig.~\ref{Fig2}(b) (dashed line). The fit yields, $\chi_0^{\rm CEF} = 1.16(1)\times 10^{-3}$~cm$^3$/mol, $\mu_{\rm eff, 0} = 2.32(2)$~$\mu_{\rm B}$, $\mu_{\rm eff, 1} = 3.97(2)$~$\mu_{\rm B}$, $\Delta^{\rm CEF}/k_{\rm B} = 92.95(1)$~K, and $\theta_{\rm CW}^{\rm CEF} = -0.8(2)$~K. Remarkably, the value of $\Delta^{\rm CEF}/k_{\rm B}$ is in good agreement with our INS results (discussed later). 

Next, we measured the magnetic isotherms ($M$ vs $H$) at four different temperatures $T = 1.85, 3, 5$, and $10$~K which are shown in Fig.~\ref{Fig2}(c). For $T = 1.85$~K, $M$ increases with $H$ and then saturates at $\mu_0H \geq 6$~T. However, even after saturation, $M$ still shows a weak increase with $H$, attributed to the Van-Vleck contribution. To estimate the Van-Vleck contribution, we performed a linear fit to $M(H)$ in the high field region ($\mu_0H \geq 6$~T). The $y$-intercept of the fit yields a saturation magnetization of $M_{\rm sat} = 1.10(1)$~$\mu_{\rm B}$/Nd$^{3+}$ while the slope gives the Van-Vleck susceptibility of $\chi_{\rm VV} = 0.0157(2)$~$\mu_{\rm B}$/T. From the value of $M_{\rm sat}$, the $g$-value is obtained to be $\sim 2.2$ considering $J_{\rm eff} = 1/2$ at low temperatures. This $g$-value is close to that obtained from the low-$T$ $\chi(T)$ analysis. 
%The electron spin resonance measurements at a lower temperature would be useful in order to precisely pin point the $g$-value.

As mentioned above, the interaction among the $J_{\rm eff} = 1/2$ spins at low temperatures is very weak and they are expected to behave like a paramagnet above $\theta_{\rm CW}^{\rm LT} = -0.26(1)$~K. Therefore, we utilized the following expression, typically used for non-interacting moments, to fit the magnetic isotherms~\cite{Sebastian034403}:
\begin{equation}
    M (H) = \chi_{\rm VV}H + N_{\rm A}g\mu_{\rm B}J_{\rm eff}B_{J_{\rm eff}} (x).
\label{BF}
\end{equation}
Here, $B_{J_{\rm eff}} (x)$ is the Brillouin function and $x = g\mu_{\rm B}J_{\rm eff}H/(k_{\rm B}T)$\cite{Kittel2004}~. For a $J_{\rm eff} = 1/2$ system, $B_{J_{\rm eff}} (x)$ turns into a simple form, $B_{J_{\rm eff}} (x) = \tanh(x)$. All the magnetic isotherms in Fig.~\ref{Fig2}(c) could be fitted well using Eq.~\eqref{BF}, suggesting uncorrelated paramagnetic spins down to 1.85~K. For this fit, we fixed the value of $\chi_{\rm VV} = 0.0157(2)$~$\mu_{\rm B}$/T and $J_{\rm eff} = 1/2$, and the fit results in a $g$-value of 2.3(1). This $g$-value is in close agreement with that obtained from low-$T$ $\chi(T)$ analysis and saturation magnetization. To further confirm the non-interacting behavior of the spins, we performed a scaling analysis of the magnetization as shown in the inset of Fig.~\ref{Fig2}(c). The Van-Vleck subtracted magnetization ($M - M_{\rm VV}$) vs $\mu_0H/T$ for all temperatures collapses onto a single curve, further endorsing a very small magnetic correlation at low temperatures.

\subsection{Heat capacity}
\begin{figure*}
	\includegraphics[width=\textwidth]{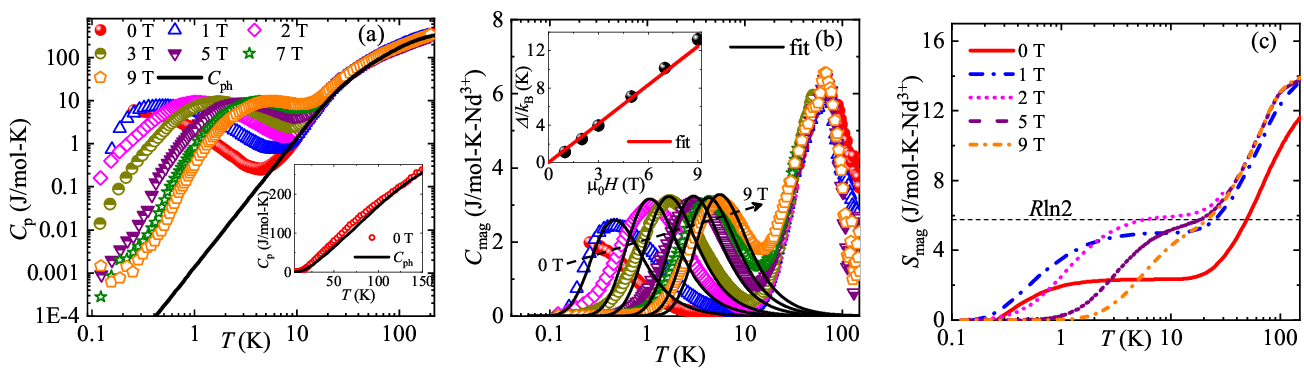}
 	\caption{\label{Fig3} (a) $C_{\rm p}(T)$ measured in different magnetic fields. The solid black line is the phonon heat capacity [$C_{\rm ph}(T)$]. Inset: Zero-field $C_{\rm p}(T)$ and $C_{\rm ph}(T)$ are shown in linear scale to highlight the feature at $\sim 68$~K. (b) $C_{\rm mag}(T)$ in various applied fields. The black solid lines are the Schottky fits as described in the text. Inset: The activated gap ($\Delta/k_{\rm B}$) vs $H$ and the red solid line is the linear fit. 
    %(Right $y$-axis) molar fraction $f$ as a function of applied field. 
    (c) Magnetic entropy $S_{\rm mag}(T)$ at some representative fields.}
 \end{figure*}
Heat capacity [$C_{\rm p}$] is an excellent thermodynamic probe for studying low temperature magnetic excitations. We have performed the heat capacity measurements down to 0.1~K in different applied magnetic fields, as depicted in Fig.~\ref{Fig3}(a). By lowering the temperature, the zero-field $C_{\rm p}(T)$ decreases systematically till 5~K. Upon lowering the temperature further, it follows an upward trend, possibly due to the emergence of magnetic short-range correlations. Surprisingly, this temperature scale is much higher than the $\theta^{\rm LT}_{\rm CW}$ value which points towards the effect of strong frustration inherent to the hyperkagome geometry. The upturn continues down to 0.1~K without exhibiting any $\lambda$-like anomaly, implying the absence of a magnetic LRO. However, under a magnetic field of $\mu_0H = 1$~T, a broad maxima appears at $\sim 0.4$~K, which shifts to higher temperatures with increasing field. This feature is often seen in multilevel systems, known as the Schottky anomaly~\cite{Guchhait214437,Mohanty134408}. In a magnetic insulator, the heat capacity comprising of two main contributions: magnetic and phonon. Due to the unavailability of a suitable non-magnetic analog, we have 
employed the linear combination of four Debye functions to extract the phonon heat capacity [$C_{\rm ph}(T)$]. The presence of four different atoms with distinct atomic masses allows us to model the heat capacity data at high temperatures using the multiple Debye functions~\cite{Ahmed214413,Nath064422}:
\begin{equation}\label{MD}
C_{\rm ph}(\theta_{\rm D},T)=9R\sum_{i =1}^{4}c_{\rm n}\left(\frac{T}{\theta_{\rm Dn}}\right)^3\int_{0}^{\theta_{\rm Dn}/T}\frac {x^4e^x}{(e^x-1)^2} \,dx \ .
\end{equation}
Here, $x=\frac{\hbar\omega}{k_{\rm B}T}$, $\omega$ is the frequency of oscillation, $R$ denotes the universal gas constant, $\theta_{\rm Dn}$ is the characteristic Debye temperature of each atom type, and $c_{\rm n}$ represents the group of different atoms in the formula unit. The high-temperature part of $C_{\rm p}(T)$ was fitted using the above expression. During the fitting, we fixed $n_1 = 3$, $n_2 = 12$, $n_3 = 3$, and $n_4 = 2$ corresponding to the Li, O, Nd, and W atoms, respectively, in the formula unit. The fit returns the following Debye temperatures: $\theta_{\rm Li}^{\rm D} = 1000(3)$~K, $\theta_{\rm O}^{\rm D} = 620(2)$~K, $\theta_{\rm Nd}^{\rm D} = 300(2)$~K, and $\theta_{\rm W}^{\rm D} = 155(1)$~K. Subsequently, the fit was extrapolated down to low temperatures and subtracted from the total heat capacity $C_{\rm p}(T)$ to obtain $C_{\rm mag}(T)$ in various applied fields, presented in Fig.~\ref{Fig3}(b).

In zero-field $C_{\rm mag}(T)$, only one broad maximum is observed at around 68~K. In contrast, under field, $C_{\rm mag}(T)$ features two broad maxima. The low temperature broad maximum corresponds to the Schottky anomaly, as discussed earlier. On the other hand, the broad maximum at higher temperatures ($\sim 68$~K) is field independent. This appears due to the transition between the ground state and higher excited state Kramers' doublets, while the former one is due to the transition between the Zeeman split ground state Kramers' doublet. Later, we will discuss these two anomalies explicitly and compare them with our CEF simulated data.

The magnetic entropy $S_{\rm mag}(T)$ is estimated by integrating $C_{\rm mag}(T)/T$ over the measured temperature range as shown in Fig.~\ref{Fig3}(c). The zero-field entropy shows a plateau at 2.4~J/mol-K at lower temperatures. This is only 41~\% of $R \ln2 \sim 5.76$~J/mol-K, implying that a large fraction of the magnetic entropy is accumulated below 100~mK, a possible signature of QSL. However, at $\mu_0H = 1$~T, the low temperature entropy shows a plateau at $\sim 5.0$~J/mol-K, and it increases further at higher temperatures. Indeed, this low-$T$ plateau value of $S_{\rm mag}(T)$ is close to $R\ln2$, suggesting that the lowest Kramers' doublet with $J_{\rm eff} = 1/2$ is the ground state. Furthermore, the value of $S_{\rm mag}(T)$ at 110~K is still smaller than the expected value for a free Nd$^{3+}$ ion with $J = 9/2$.

%However, in the present study in zero field, only a small fraction of entropy is released, implying that a large fraction of the entropy is accumulated below 2~K due to the buildup of magnetic correlations at lower temperatures. With the application of magnetic field, the magnetic entropy shifts towards high temperatures, and we are able to recover nearly 90~\% of $Rln2$ at $\mu_0H = 5$~T. This further supports a Kramers' doublet with a $J_{\rm eff} = 1/2$ ground state at low temperatures.

To quantify the Schottky contribution, we fitted $C_{\rm mag}(T)$ using the two-level Schottky function~\cite{Kittel2004},
\begin{equation}\label{Schottky}
    C_{\rm Sch} (T,H) = fR\left(\frac{\Delta}{k_{\rm B}T}\right)^2 \frac{e^\frac{\Delta}{k_BT}}{\left[e^\frac{\Delta}{k_BT}+1\right]^2}.
\end{equation}
Here, $f$ stands for the molar fraction of the free spins and $\Delta/k_{\rm B}$ is the energy gap between the Zeeman-split ground state doublet. Figure~\ref{Fig3}(b) displays the Schottky fits (solid black line) using Eq.~\eqref{Schottky}. The $C_{\rm mag}(T)$ curves at all the fields are well described by this model and the obtained field variation of $\Delta/k_{\rm B}$ is shown in the inset of Fig.~\ref{Fig3}(b). 
%As the field increases, $f$ increases and then saturates to a value $\sim 0.70$ at higher fields, which signifies $\sim 70$~\% of free Nd$^{3+}$ spins are excited to the higher-energy levels and contribute to the Schottky anomaly. 
The energy gap $\Delta/k_{\rm B}$ follows a linear behavior with field. Using the value of $\Delta/k_{\rm B}\simeq 13.2$~K at 9~T in $\Delta/k_{\rm B} = g\mu_{\rm B}H/k_{\rm B}$, the $g$-value is estimated to be $2.2(2)$, which is closely aligned with the value obtained from the magnetization analysis. This further confirms that the observed Schottky effect in the heat capacity arises from the ground-state Kramers' doublet with $J_{\rm eff}=1/2$.

%A scaling plot of $C_{\rm mag}$ vs $k_{\rm B}T/(M_{\rm sat}\mu_0H)$ for different fields is presented in the inset of Fig.~\ref{Fig3}(c). Notably, the data for fields $\mu_0H\leq 2$~T collapse onto a single curve. This reflects that the interaction strength between Yb$^{3+}$ ions is weak, further supporting the minuscule value of $\theta_{\rm CW}^{\rm LT}$.

\subsection{Inelastic Neutron Scattering}
%In this section, we discuss the INS measurement carried out on a high-quality polycrystalline Nd$_3$Li$_3$W$_2$O$_{12}$ sample. Since the magnetic properties of Nd$_3$Li$_3$W$_2$O$_{12}$ depends on the CEF scheme of the Nd$^{3+}$ ion, 
%we probed the CEF excitations via the INS experiments. Figure~\ref{Fig4} shows the two-dimensional (2D) contour plot of INS spectra at different temperatures (6, 100, and 250~K) for multiple incident energies ($E_i =$ 29.7, 60, and 180~meV)~\cite{INS_Data_Ref}. 
%We observed three non-dispersive excitations at around 9.5, 22, and 25.5~meV for $T = 6$~K and $E_i = 60$~meV [see Fig.~\ref{Fig4}(a)]. The spectrum at $T = 6$~K for high incident energy, $E_i = 180$~meV, also shows a weak excitation at around 90~meV 
%as shown in Fig.~\ref{Fig4}(b). Thus, we observed four well-defined CEF excitations at $T = 6$~K in Nd$_3$Li$_3$W$_2$O$_{12}$, as expected for a Nd$^{3+}$ based system, corresponding to the transitions from ground state doublet to four high-energy doublets. 
%Furthermore, INS spectra at 100 and 250~K [see Figs.~\ref{Fig4}(c) and (d)] reveal several additional weak excitations in the low-$Q$ regime, reflecting transitions between higher-lying CEF levels.

\begin{figure*}
	\includegraphics[scale=0.5]{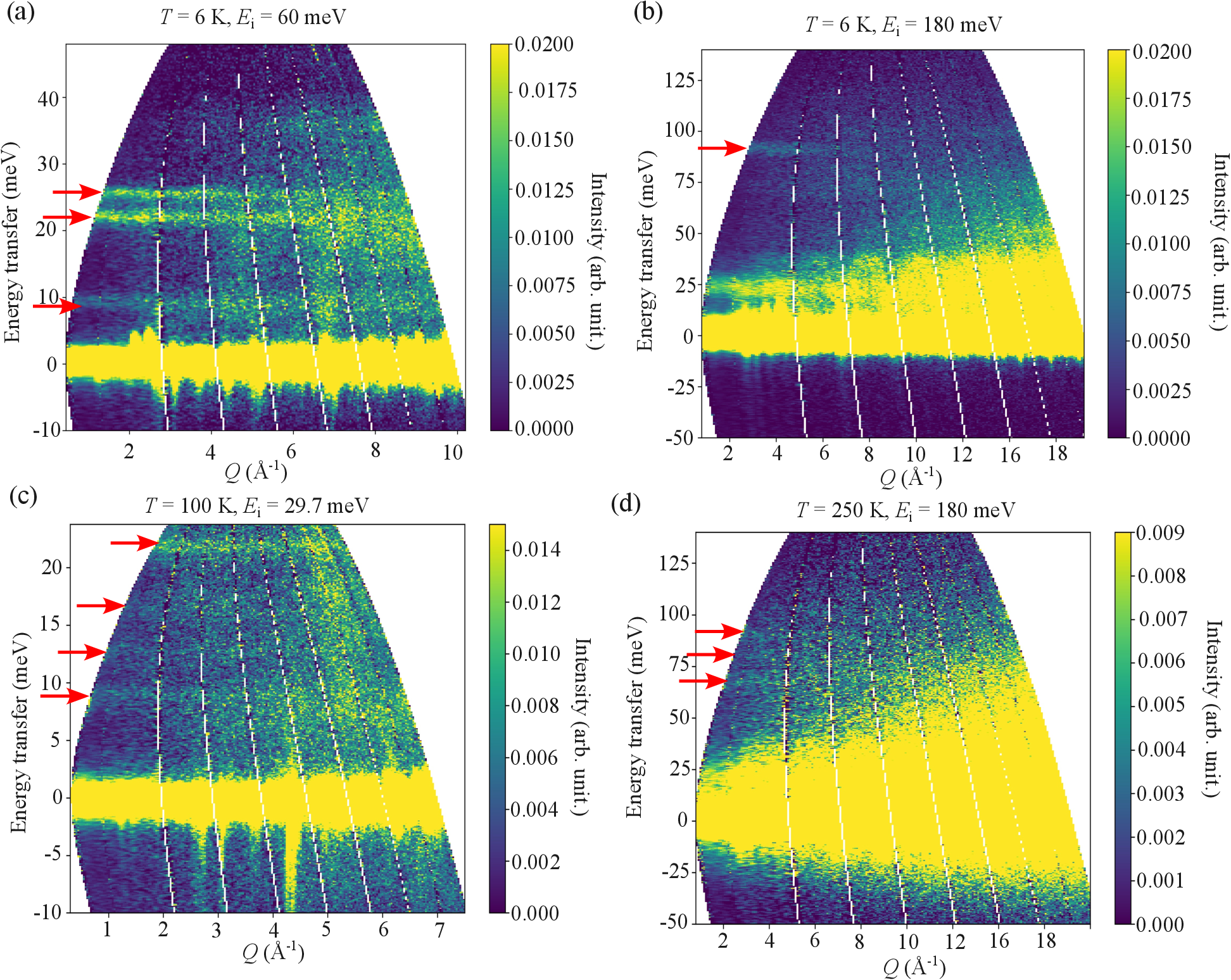}
	\caption{(a-d) Raw INS spectra of Nd$_3$Li$_3$W$_2$O$_{12}$ at different temperatures ($T = 6$, 100, and 250~K) and in different incident neutron energies ($E_i = 29.7$, 60, and 180~meV). The red arrows mark the CEF excitations.}
	\label{Fig4}
\end{figure*}

In order to directly assess the influence of Nd$^{3+}$ crystal field excitations on the magnetic properties of Nd$_3$Li$_3$W$_2$O$_{12}$, we carried out a zero-field INS measurement on a high-quality polycrystalline sample.
Figure~\ref{Fig4} shows two-dimensional (2D) contour plots of INS spectra obtained at different temperatures (6, 100, and 250~K) for multiple incident energies ($E_i =$ 29.7, 60, and 180~meV)~\cite{INS_Data_Ref}. 
The neutron scattering cross-sections for nuclear and magnetic scattering imply that CEF excitations are expected to be most intense at low $Q$ while phonon scattering become dominant at high $Q$~\cite{Boothroyd2020}. The latter effect is particularly clear from the high-$E_i$ data sets. We observe three non-dispersive excitations around 9.5, 22, and 25.5~meV for $T = 6$~K and $E_i = 60$~meV [see Fig.~\ref{Fig4}(a)]. The spectrum at $T = 6$~K for high incident energy, $E_i = 180$~meV, contains a weak but clear excitation around 90~meV for small values of $Q$ as shown in Fig.~\ref{Fig4}(b). Thus, at our experimental base temperature we observe four well-defined CEF excitations as expected for a Nd$^{3+}$ based system and corresponding to the transitions from the ground state Kramers doublet to four higher-energy doublets (I $\to$ II, III, IV, and V). Furthermore, INS spectra at 100 and 250~K [see Figs.~\ref{Fig4}(c) and (d)] reveal several additional weak excitations in the low-$Q$ regime, reflecting transitions between higher-lying CEF levels as these become thermally occupied.

\begin{figure*}
	\includegraphics[scale=2.3]{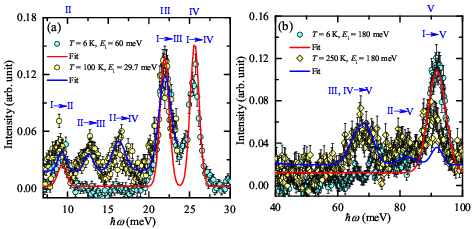}
	\caption{The INS spectral intensity (after subtraction of the phonon contribution discussed in the main text) as a function of energy transfer (for $T= 6$, 100, 250~K and $E_{\rm i}=$ 29.7, 60, 180~meV) obtained by integrating intensity in the low-$Q$ regime. The solid line is the corresponding fit using the CEF Hamiltonian. The energies of the four excited doublets are shown (II to V) on the top. Each peak is labeled with the corresponding transition. The $E_i=29.7$ meV data are displaced vertically by 0.01 intensity units.}
	\label{Fig5}
\end{figure*}

%To clearly visualize all the CEF excitations and facilitate quantitative analysis of the INS spectra, the scattering intensity was integrated over the low-$Q$ (0~$\leq Q \leq$~3~\AA$^{-1}$) and high-$Q$ regions to obtain the energy-transfer dependence of the INS intensity. Since no nonmagnetic analog compound is available, the phonon background was estimated from the high-$Q$ data, where phonon scattering dominates, and appropriately scaled before being subtracted from the low-$Q$ spectra.
To quantitatively analyse the INS spectra, the raw scattering intensity was integrated over the low-$Q$ (0~$\leq Q \leq$~3~\AA$^{-1}$) and high-$Q$ regions (5.5~$\leq Q \leq$~7.5~\AA$^{-1}$, 6.5~$\leq Q \leq$~9.5~\AA$^{-1}$ and 12~$\leq Q \leq$~18~\AA$^{-1}$ for $E_i=29.7$, $60$ and $180$ meV, respectively) 
to obtain the energy-transfer dependence of the intensity. 
In the absence of a nonmagnetic analog for Nd$_3$Li$_3$W$_2$O$_{12}$, the purely magnetic scattering was estimated by subtracting the phonon-dominated high-$Q$ data from the low-$Q$ data after appropriate scaling at energy transfers where no CEF modes are observed.
%The resulting phonon-subtracted spectra were subsequently used to identify the CEF excitations and perform the CEF analysis.
%In order to clearly visualize all the CEF modes and fit the INS data, we have subtracted the scaled phonon excitation which dominant in high-$Q$ region from the low-$Q$ (0~$\leq Q \leq$~3~\AA$^{-1}$) INS data, since no non-magnetic analog compound is available. 
The resulting intensity versus energy transfer ($\hbar \omega$) plots are presented in Fig.~\ref{Fig5} for $T = 6$, $100$, and $250$~K with $E_i = 29.7$, $60$, and $180$~meV.
%The strong signal at $\hbar \omega = 0$~meV corresponds to quasielastic neutron scattering. 
For $T = 6$~K, the four CEF excitations mentioned above appear as well-defined peaks around 9.5 (II), 22 (III), 25.5 (IV), and 90 (V)~meV.
%All the transitions from the ground-state doublet (I) to the excited CEF levels (I $\to$ II, III, IV, and V) lie below 100~meV and are fully resolved. At low temperatures, the observed excitations occur from the Kramers doublet ground state to one of the excited state doublets (THE PRECEEDING TWO SENTENCES DO NOT APPEAR TO ADD ANYTHING NEW).
At elevated temperatures, thermal population of the 
low-lying excited CEF levels (primarily II, III, and IV) becomes significant. In the 100~K spectrum, two weak excitations around 12.7 and 16.4~meV [see Fig.~\ref{Fig5}(a)] are detected. These can be associated with transition between the first (II) excited Kramers doublet and the third (III) and fourth (IV) doublets (II $\to$ III and II $\to$ IV), respectively. Similarly, the 250~K spectrum for $E_i = 180$~meV shows two additional peaks around 68.4~meV and 80~meV, as shown in Fig.~\ref{Fig5}(b). The broad 68.4~meV peak corresponds to the combination of transitions from the second (III) and third (IV) Kramers doublet to the fourth (V) doublet (III, IV $\to$ V), whereas the 80~meV peak appears due to the transition between the II and V doublets (II $\to$ V). Although our model also predicts a III $\to$ IV transition, we did not observe it in the experimental data because of the very weak transition probability.

\subsection{CEF analysis}
The INS intensity versus energy transfer data can be analyzed using an appropriate CEF Hamiltonian for this compound. Following the Stevens formalism, the CEF Hamiltonian can be written as $\mathcal{H}_{\rm CEF} = \sum_{l,m}B_l^m\hat{O}_l^m$~\cite{Stevens209}.
%\begin{equation}\label{CEF}
%	\mathcal{H}_{\rm CEF} = \sum_{l,m}B_l^m\hat{O}_l^m.
%\end{equation}
In this expression, $\hat{O}^m_l$ are the standard Stevens operators~\cite{Huthings227,Stevens209}, which are formulated in terms of the angular momentum operators~\cite{Newman2000}. The coefficients $B^m_l$ are referred as the CEF parameters, associated with the electronic structure of the rare-earth materials~\cite{Guchhait144434,Huthings227}. For $f$-electron systems, $l$ takes even integer values from 0 to 6, while $m$ varies between $-l$ and $l$. In Nd$_3$Li$_3$W$_2$O$_{12}$, Nd$^{3+}$ ions occupy sites with  $D_{2}$ point group symmetry and consequently, the CEF model Hamiltonian for this compound can be represented as:
\begin{align}
	\label{CEF1}
	\begin{split}
	&\mathcal{H}_{\rm CEF} = B_2^0\hat{O}_2^0 + B_2^2\hat{O}_2^2 + B_4^0\hat{O}_4^0 + B_4^2\hat{O}_4^2 \\
	& + B_4^4\hat{O}_4^4 + B_6^0\hat{O}_6^0 + B_6^2\hat{O}_6^2 + B_6^4\hat{O}_6^4 + B_6^6\hat{O}_6^6 .
   \end{split}
\end{align}

\begin{figure*}
	\includegraphics[width=\textwidth]{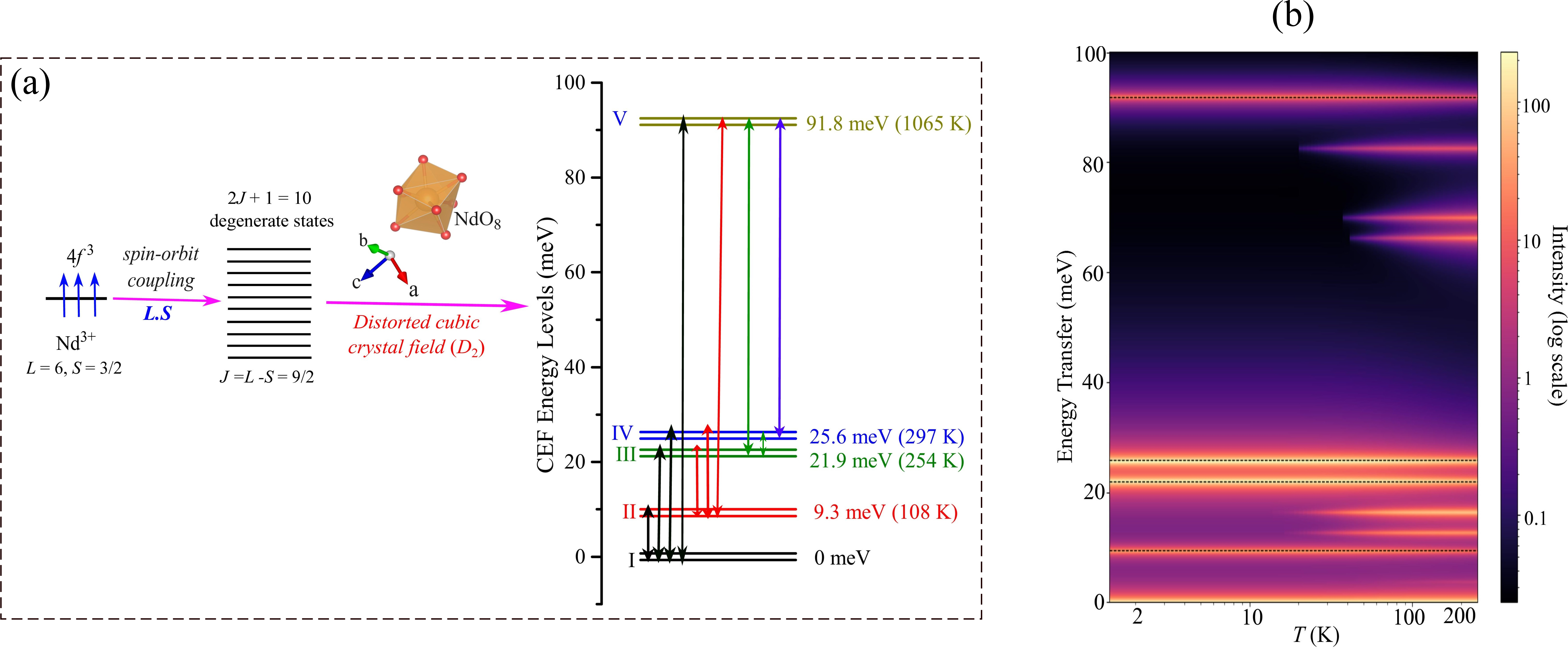}
	\caption{\label{Fig6} (a) Schematic representation of CEF energy level scheme obtained from the zero-field INS data. A distorted NdO$_8$ polyhedron formed by Nd$^{3+}$ and O$^{2-}$ ions is shown, that generates the CEF potential. (b) A contour plot of the calculated CEF neutron-scattering intensity as a function of temperature and energy transfer. The observed bands correspond to allowed transitions depicted in Fig.~\ref{Fig6}(a) in different temperature regimes.}
\end{figure*}

As illustrated in Figs.~\ref{Fig5}, we fitted the 6, 100, and 250~K data simultaneously employing the above CEF Hamiltonian within the Mantid software framework~\cite{Arnold156}. 
We obtained 9 distinct sets of CEF parameters that reproduce the INS data equally well.
%and yield similar powder magnetic susceptibility and magnetization consistent with the experimental data. 
To further constrain the solution, the calculated powder magnetic susceptibility [$\chi_{\rm CEF}(T,H)$], magnetization [$M_{\rm CEF}(T,H)$], and heat capacity [$C_{\rm CEF}(T,H)$] data, corresponding to each parameter set were compared with the experimental data. The parameter set exhibiting the best overall agreement with the experimental data was eventually selected. The obtained final set of CEF parameters are listed in Table~\ref{CEF_Para}. Then, the CEF Hamiltonian was diagonalized to determine the CEF energy eigenvalues of the compound. The obtained energy eigenvalues are 0, 9.3, 21.9, 25.6, and 91.8~meV, corresponding to five Kramers doublets, as depicted in Fig.~\ref{Fig6}. The eigen functions associated with the Kramers doublets can be expressed as linear superpositions of the basis states, $|\psi_k,\pm\rangle=\sum_{m_J = -9/2}^{m_J = 9/2}C_{m_J}^{k,\pm}\left|J = 9/2, m_J \right\rangle$.
%\begin{equation}\label{CEF_wave_vector}
%	|\psi_k,\pm\rangle=\sum_{m_J = -15/2}^{m_J = 15/2}C_{m_J}^{k,\pm}\left|J = 15/2, m_J \right\rangle.
%\end{equation}
Here, $C_{m_J}^{k,\pm}$ are the weighted coefficients that quantify the contribution of each ($\left|J=9/2,m_J\right\rangle$) basis state to the corresponding eigenfunction. The complete set of CEF energy eigenvalues, together with the associated coefficients ($C_{m_J}^{k,\pm}$) for the various eigenstates of Nd$_3$Li$_3$W$_2$O$_{12}$, is summarized in Table~\ref{Eigenvalue_and_Eigervector}. From the table, the wave function corresponding to the ground-state doublet can be written as 
\begin{align}\label{Wavefunction}
|\psi_0,\pm\rangle & =\mp0.3421\left|\mp\frac{9}{2}\right\rangle -0.0177\left|\mp\frac{7}{2}\right\rangle\mp0.2291\left|\mp\frac{5}{2}\right\rangle \nonumber\\
&\quad -0.0037\left|\mp\frac{3}{2}\right\rangle \pm0.5607\left|\mp\frac{1}{2}\right\rangle +0.0142\left|\pm\frac{1}{2}\right\rangle \nonumber\\
&\quad \mp0.1473\left|\pm\frac{3}{2}\right\rangle - 0.0058\left|\pm\frac{5}{2}\right\rangle \mp0.7027\left|\pm\frac{7}{2}\right\rangle \nonumber\\
&\quad -0.0086\left|\pm\frac{9}{2}\right\rangle. 
\end{align}
This ground state wavefunction is an admixture of $m_J$ values. The non-zero value of $\langle \psi_0, +| J_{\pm}|\psi_0,-\rangle$ implies a significant overlap between the ground state Kramers’ doublet, which can facilitate quantum tunneling among these states. This makes the ground state highly quantum in nature, similar to several other rare-earth-based magnets~\cite{Scheie144432,Guchhait144434,Sebastian104428,Guchhait214437}.

Using the above wave-functions of the CEF ground state ($|\psi_0,\pm\rangle $), one can evaluate the anisotropic $g$-tensor components, using the expression $g_{(\alpha ~ = ~x, ~y, ~z)} = 2g_{J}\langle\psi_0,\pm|J_{(\alpha ~ = ~x, ~y, ~z)}|\psi_0,\pm\rangle$~\cite{Kutuzov012039}. The calculated value of the $g$-components are $g_{x} = 2.66 (1)$, $g_{y} = 1.99(1)$, and $g_{z} = 1.32(1)$, respectively. The average value of $g$-factor [$g_{\rm avg} = \sqrt{(g_{x}^2 + g_{y}^2 + g_{z}^2)/3}$] is found to be $g_{\rm avg} = 2.1(2)$, which is in good agreement with the value inferred from the low temperature $\chi(T)$ analysis and the saturation magnetization $M_{\rm sat}$. Likewise, the wave functions of the excited CEF doublets can be constructed using the affiliated coefficients listed in Table~\ref{Eigenvalue_and_Eigervector}, enabling the determination of the $g$-tensor components for the higher-lying crystal-field levels.
\begin{table}[ptb]
	\caption{Fitted CEF parameters for Nd$_3$Li$_3$W$_2$O$_{12}$.}
	\label{CEF_Para}
	\begin{ruledtabular}
		\begin{tabular}{cccc}
			$B_l^m$ (meV) & Values & $B_l^m$ (meV) & Values \\\hline
			$B_2^0$ & -0.1422 & $B_6^0$ & 0.0003\\
			$B_2^2$ & -1.0851 & $B_6^2$ & -0.0002\\
			$B_4^0$ & -0.0052 & $B_6^4$ & -0.0011\\
			$B_4^2$ & 0.0302 & $B_6^6$ & -0.0041\\
			$B_4^4$ & 0.0689 \\
		\end{tabular}
	\end{ruledtabular}
\end{table}

Figure~\ref{Fig6} depicts the CEF energy-level scheme derived from the INS data, where the spin-orbit-entangled multiplet is split into five doublets by the $D_2$-symmetric crystal field environment. The arrows indicate the allowed INS transitions between the CEF levels. The rightmost panel shows a contour plot of the INS spectra simulated using the optimized CEF parameters as a function of temperature (up to 250~K) and energy transfer (up to 100~meV), illustrating the evolution of the CEF excitations with increasing temperature. Four intense excitations, highlighted by dashed lines, persist throughout the entire temperature range due to transitions from the ground-state doublet to the excited state doublets. As the temperature increases, additional spectral features emerge related to transitions between the thermally populated excited CEF levels as discussed above. Here, at around 3~meV, above 100~K, we observed a weak spectral line, which indicates a very small transition probability between III and IV levels, consistent with the observed INS spectra.

\begin{figure*}
	\includegraphics[width=\textwidth]{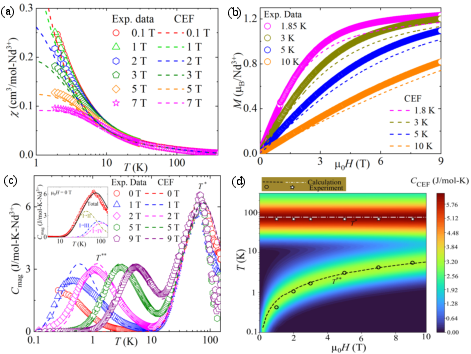}
	\caption{\label{Fig7} (a) Experimental $\chi(T)$ data together with the calculated susceptibility from the CEF model. (b) Isothermal magnetization curves simulated using the CEF model at various temperatures and compared with the corresponding experimental results. (c) Calculated CEF contribution to the heat capacity, [$C_{\rm CEF}(T)$], under different fields, alongside the experimentally determined magnetic heat capacity, ($C{\rm mag}(T)$). Inset: Zero-field $C_{\rm mag}(T)$ with simulated CEF data (dash-dotted line) for the transition from ground state to the higher excited state doublets. The black solid line is the summation for all the three transitions.
    (d) 2D contour map of the $C_{\rm CEF}$ as a function of field and temperature. On the top of this plot, $T^{\star}$, and $T^{\star\star}$ obtained from the experimental $C_{\rm mag}$ (symbols) and calculated $C_{\rm CEF}$ (dotted lines) [from Fig.~\ref{Fig7}(c)] are shown.}
\end{figure*}
To assess the effect of CEF excitations on the physical properties, we calculated the contribution of the crystal field excitations to the magnetic susceptibility [$\chi_{\rm CEF}(T)$], magnetization isotherms [$M_{\rm CEF}(H)$], and heat capacity [$C_{\rm CEF}(T)$] using the calculated CEF eigenenergies and eigenfunctions, incorporating the Zeeman splitting induced by an applied magnetic field. The methodology employed for these calculations is described in Appendix~B. As evident from Figs.~\ref{Fig7}(a) and \ref{Fig7}(b) the calculated $\chi_{\rm CEF}(T)$ and $M_{\rm CEF}(H)$ reproduce the experimental data. The slight deviation observed between the experimental and computed results at low temperatures are likely attributable to the weak magnetic exchange coupling between the Nd$^{3+}$ ions in Nd$_3$Li$_3$W$_2$O$_{12}$, which is not taken into account for the calculations.
\begin{table*}
\caption{Energy eigenvalues and the coefficients ($C_{m_J}^{k,\pm}$) corresponding to different eigenstates of the CEF Hamiltonian for Nd$_3$Li$_3$W$_2$O$_{12}$.}
\label{Eigenvalue_and_Eigervector}
\begin{ruledtabular}
\begin{tabular}{c|cccccccccc}
$E$ (meV) &$|-\frac{9}{2}\rangle$ & $|-\frac{7}{2}\rangle$ & $|-\frac{5}{2}\rangle$ & $| -\frac{3}{2}\rangle$ & $|-\frac{1}{2}\rangle$ & $|\frac{1}{2}\rangle$ & $|\frac{3}{2}\rangle$ & $|\frac{5}{2}\rangle$ & $|\frac{7}{2}\rangle$ & $|\frac{9}{2}\rangle$ \tabularnewline
 \hline 
0.00 & -0.3421 & -0.0177 & -0.2291 & -0.0037 & 0.5607 & 0.0142 & -0.1473 & -0.0058 & -0.7027  & -0.0086 \tabularnewline
0.00 & -0.0086 & 0.7027 & -0.0058 & 0.1473 & 0.0142 & -0.5607 & -0.0037 & 0.2291 & -0.0177 & 0.3421 \tabularnewline
9.31 & 0.6423 & -0.0083 & 0.3446 & -0.0271 & 0.6422 & -0.1025 & 0.1695 & -0.0550 & 0.0518 & -0.1026 \tabularnewline
9.31 & 0.1026 & 0.0518 & 0.0550 & 0.1695 & 0.1025 & 0.6422 & 0.0271 & 0.3446 & 0.0083 & 0.6423 \tabularnewline
21.94 & -0.0079 & 0.0321 & 0.0454 & -0.8064 & 0.0002 & 0.0022 & -0.0636 & 0.5764 & 0.0025 & -0.1012 \tabularnewline
21.94 & 0.1012 & 0.0025 & -0.5764 & -0.0636 & -0.0022 & 0.0002 & 0.8064 & 0.0454 & -0.0321 & -0.0079 \tabularnewline
25.62 & -0.0048 & 0.6173 & 0.0006 & 0.1532 & 0.0038 & 0.4749 & 0.0012 & 0.0721 & 0.0049 & -0.6038 \tabularnewline
25.62 & 0.6038 & 0.0049 & -0.0721 & 0.0012 & -0.4749 & 0.0038 & -0.1532 & 0.0006 & -0.6173 & -0.0048 \tabularnewline
91.8 & 0.0000 & 0.3479 & 0.0000 & -0.5205 & 0.0000 & 0.1920 & 0.0000 & -0.6973 & 0.0000 & 0.2914 \tabularnewline
91.8 & -0.2914 & 0.0000 & 0.6973 & 0.0000 & -0.1920 & 0.0000 & 0.5205 & 0.0000 & -0.3479 & 0.0000 \tabularnewline
\end{tabular}\end{ruledtabular}
\end{table*}

Figure~\ref{Fig7}(c) presents $C_{\rm CEF}(T)$ under different fields. In zero-field, the $C_{\rm CEF}(T)$ exhibits a broad maximum at $T^{*} \simeq 68$~K, consistent with the experimental $C_{\rm mag}(T)$. This feature appears due to the thermal excitations from the ground state doublet to the first, second, and third excited doublets [see inset of Fig.~\ref{Fig7}(c)]. The calculated $C_{\rm CEF}$ approaches zero below about 10~K, unlike the low temperature upturn observed in the experimental $C_{\rm mag}(T)$ data. This upturn reflects the buildup of short-range magnetic correlations between Nd$^{3+}$ ions. In an applied magnetic field, the degenerate Kramers' doublets split further, and the calculated $C_{\rm CEF}$ results in another low-$T$ broad peak reproducing our experimental $C_{\rm mag}(T)$. At $\mu_0 H = 1$~T, the low-$T$ maximum appears near $T^{**}\sim 0.4$~K, which can be attributed to the transition occurring between Zeeman levels of the ground-state doublet. With increasing field, this peak broadens and shifts to the higher temperatures, consistent with the experimental $C_{\rm mag}(T)$ curve. On the other hand, the high temperature maximum at $T^{*}$ remains nearly unchanged in position and exhibits only slight broadening, in agreement with the experimental $C_{\rm mag}(T)$ data.
%\textbf{This upturn reflects the buildup of short-range magnetic correlations between Nd$^{3+}$, since the value of $\theta_{\rm CW}^{\rm LT}$ is of the same order of magnitude.}

For a comparison between the calculated and experimental data, we constructed a 2D contour map of $C_{\rm CEF}(T,H)$ along with the characteristic temperatures $T^{\star}$ and $T^{\star\star}$ obtained from the experimental $C_{\rm mag}(T)$ [from Fig.~\ref{Fig3}(b)], as depicted in Fig.~\ref{Fig7}(d). The magnetic-field dependence of these anomalies extracted from the calculations follows the trend of experimental $C_{\rm mag}(T)$ data, further validating the CEF model of Nd$_3$Li$_3$W$_2$O$_{12}$.
%The minor deviation between the experimental and theoretical values of $T^{\star}$ and $T^{\star\star}$ may arise from a weak anisotropic exchange interactions among the Nd$^{3+}$ ions, which is not incorporated into the present calculations.

\section{Discussion and Summary}
We have investigated the ground-state properties of the previously unexplored Nd$^{3+}$-based 3D frustrated hyperkagome system Nd$_3$Li$_3$W$_2$O$_{12}$ through bulk magnetic measurements and INS experiments. The small value of $\theta_{\rm CW}^{\rm LT} = -0.26(1)$~K indicates a weak AFM interaction among the Nd$^{3+}$ moments. The dipolar interaction between the nearest-neighbour (NN) Nd$^{3+}$ ions (with a distance $d_{\rm NN} \simeq 3.817$~\AA) of this compound is estimated to be $E_{\rm dip} \simeq \frac{\mu_0g_{\rm avg}^2\mu_{\rm B}^2 J_{\rm eff}^2}{4\pi d^3}  = 0.015(1)$~K~\cite{Xiang2023}, where, $\mu_0$ is the vacuum permeability, $J_{\rm eff} = 1/2$, and $g_{\rm avg} = 2.3(1)$. This strength of dipolar coupling is approximately an order of magnitude smaller than $\theta_{\rm CW}^{\rm LT}$, suggesting that the low temperature magnetic behavior is expected to be governed primarily by exchange interactions rather than dipolar coupling. No signature of the magnetic LRO is observed down to 0.1~K, reflecting the role of geometric frustration in suppressing the magnetic ordering.

The low temperature magnetization, heat capacity, and the INS data further imply a pseudo-spin-$1/2$ ($J_{\rm eff}=1/2$) ground state for this compound. Typically, for the systems with $J_{\rm eff}=1/2$ ground state doublet, the ratio of the squared-moments $R\equiv \left(\frac{\mu_{\rm eff}}{\mu_{\rm sat}}\right)^2$ is expected to approach a value $R=3$. For Nd$_3$Li$_3$W$_2$O$_{12}$, we obtained $R_{\rm exp} = 3.5(4)$ in the low-$T$ regime, which confirms that the lowest Kramers' doublet with $J_{\rm eff}=1/2$ is the ground state~\cite{Guo094404}. The CEF analysis of INS data reveals that the first excited doublet lies at an energy of approximately 9~meV ($> 100$~K) above the ground state doublet. Consequently, at temperatures well below this energy scale, the magnetic properties are governed exclusively by the lowest Kramers' doublet with $J_{\rm eff} = 1/2$ ground state of the compound. This conclusion is further supported by the recovery of a magnetic entropy to $S_{\rm mag} \simeq R\ln2$ at low temperatures.

%\section{Summary}
In summary, we report a detailed experimental investigation of the Nd$^{3+}$-based frustrated hyperkagome compound Nd$_3$Li$_3$W$_2$O$_{12}$ using magnetization, heat capacity, and INS measurements. No evidence of magnetic LRO is observed down to 0.1~K. It shows the development of AFM correlations below $\sim 2$~K in zero field, despite a small negative value of $\theta_{\rm CW}^{\rm LT}$, implying magnetic frustration in the compound. Zero-field heat capacity data manifest a single broad maximum centered at $T^* \simeq 68$~K arising from multiple CEF excitations, together with a low temperature upturn associated with the development of magnetic correlations. Application of a magnetic field suppresses the weak magnetic correlations and induces an additional broad anomaly or Schottky anomaly at low temperatures ($T^{**}$), attributable to excitations between the Zeeman-split levels of the low-energy CEF doublets. Additionally, the INS measurements directly resolve the CEF excitation spectrum, allowing the determination of the CEF parameters through simultaneous fitting of the observed excitation energies and scattering intensities at different temperatures.
  %The ground state doublet has a significant $J_z= \pm /$ component in the wave function which indicates the strong quantum effects in this compound at low temperatures. 
The INS measurements establish a CEF energy gap of approximately 100~K ($\sim 9$~meV) between the ground-state and first excited Kramers' doublets, which supports $J_{\rm eff} = 1/2$ ground state at low temperatures, consistent with the conclusions drawn from the magnetization and heat-capacity analysis at low temperatures.

Finally, using the CEF eigenenergies and eigenfunctions determined from the INS analysis, we computed $\chi_{\rm CEF}(T,H)$, $M_{\rm CEF}(T, H)$, and $C_{\rm CEF}(T, H)$, which reproduce the experimental data nicely. Furthermore, the calculated temperature-dependent CEF excitation spectrum up to 250~K successfully reproduces the evolution of the INS intensity map as a function of energy transfer and temperature. Since no magnetic LRO is detected down to 0.1 K, further measurements at lower temperatures are required to confirm whether Nd$_3$Li$_3$W$_2$O$_{12}$ ultimately undergoes a magnetic LRO or realizes a more exotic quantum-disordered ground state, such as a QSL, in the zero-temperature limit.

\section{acknowledgments}
RK and RN would like to acknowledge SERB, India, for financial support bearing sanction Grant No. CRG/2022/000997. SG and NBC gratefully acknowledge the support of the Danish Agency for Science, Technology, and Innovation through the instrument center Danscatt and of the Danish National Committee for Research Infrastructure (NUFI) through the ESS-Lighthouse Q-MAT.
%SG and NBC were supported by the Danish National Committee for Research Infrastructure (NUFI) through the ESS-Lighthouse Q-MAT and by the Danish Agency for Science, Technology, and Innovation through the instrument centre Danscatt. 
Experiments at the ISIS Neutron and Muon Source were supported by a beamtime allocation RB2610598 from the Science and Technology Facilities Council.

\section{APPENDIX A:}
Steven operators in Eq.~\eqref{CEF1} can be expressed in terms of angular momentum operators $J_{+}$ (raising operator), $J_{-}$ (lowering operator), and $J_{z}$ as~\cite{Newman2000}
\begin{align}\label{Steven}
\hat{O}_2^0&=[3J_z^2-X],\notag\\
\hat{O}_2^2&=\frac{1}{2}[J_+^2 + J_-^2] = [J_x^2 - J_y^2],\notag\\
\hat{O}_4^0&=[35J_z^4-(30X-25)J_z^2+3X^2-6X],\notag\\
\hat{O}_4^2&=\frac{1}{4}[(J_{+}^2+J_{-}^2)(7J_z^2-X-5)+ (7J_z^2-X-5)(J_{+}^2+J_{-}^2)],\notag\\
\hat{O}_4^4&=\frac{1}{2}[J_{+}^4+J_{-}^4],\notag\\
\hat{O}_6^0&=[231J_z^6-(315X-735)J_z^4+(105X^2-525X+294)J_z^2\notag\\&-5X^3+40X^2-60X],\notag\\
\hat{O}_6^2&=\frac{1}{4}[(J_{+}^2+J_{-}^2)(33J_z^4-(18X+123)J_z^2 + X^2 + 10X \notag\\& + 102)+(33J_z^4-(18X+123)J_z^2 + X^2 + 10X \notag\\& + 102)(J_{+}^2+J_{-}^2)],\notag\\
\hat{O}_6^4&=\frac{1}{4}[(J_{+}^4+J_{-}^4)(11J_z^2-X-38)+\notag \\&(11J_z^2-X-38)(J_{+}^4+J_{-}^4)],\notag\\
\hat{O}_6^6&=\frac{1}{2}[J_{+}^6+J_{-}^6].\notag\\
\end{align}
Here, $X = J(J + 1)$. 

\section{APPENDIX B:}
The magnetization is estimated by calculating the expectation value of total angular momentum ($\hat{J}$) with components $J_x$, $J_y$, and $J_z$ as~\cite{Guchhait214437}
\begin{equation}
    \begin{split}
M_{\rm CEF}(T,H)= \frac{N_{\rm A}\,g\,\mu_{\rm B}}{Z}\times
\\\sum_k e^{-\frac{E_k(H)}{k_{\rm B}T}}
\langle \psi_k(H) \lvert \hat{J}_{\alpha~=~x,~y,~z} \rvert \psi_k(H)\rangle.
	\label{M_CEF} 
\end{split}
\end{equation}
Here, $Z = \sum_ke^{-E_k(H)/k_{\rm B}T}$ is the partition function, where the summation is taken over all the energy states. $|\psi_k(H)\rangle$ and $E_k(H)$ are the $k^{\rm th}$ eigenstate and eigenvalue of the effective Hamiltonian $\mathcal{H}_{\rm eff} = \mathcal{H}_{\rm CEF} + g\mu_{\rm B}\vec{B}.\vec{J}$, with an applied field $\vec{B}$. $\chi_{\rm CEF}(T, H)$ can be calculated by taking the first derivative of $M_{\rm CEF}(T,H)$ with respect to $H$. Heat capacity [$C_{\rm CEF}(T, H$)] for a $N$ level system can be expressed as:
\begin{equation}
\begin{split}
  C_{\rm CEF}(T, H) = \frac{R}{(Zk_{\rm B}T)^2} \sum_{n>m}^{N}[E_n(H) - E_m(H)]^2 \\ \times exp\left[-\frac{E_n(H) + E_m(H)}{k_{\rm B}T}\right],
 \label{C_CEF}
 \end{split}
\end{equation}
where, $R$ is the universal gas constant. Here, $E_n$ and $E_m$ are the energy of the $n^{th}$ and $m^{th}$ CEF levels, respectively~\cite{Guchhait144434,Guchhait214437}.

%\bibliography{Ref}

\begin{thebibliography}{56}%
	\makeatletter
	\providecommand \@ifxundefined [1]{%
		\@ifx{#1\undefined}
	}%
	\providecommand \@ifnum [1]{%
		\ifnum #1\expandafter \@firstoftwo
		\else \expandafter \@secondoftwo
		\fi
	}%
	\providecommand \@ifx [1]{%
		\ifx #1\expandafter \@firstoftwo
		\else \expandafter \@secondoftwo
		\fi
	}%
	\providecommand \natexlab [1]{#1}%
	\providecommand \enquote  [1]{``#1''}%
	\providecommand \bibnamefont  [1]{#1}%
	\providecommand \bibfnamefont [1]{#1}%
	\providecommand \citenamefont [1]{#1}%
	\providecommand \href@noop [0]{\@secondoftwo}%
	\providecommand \href [0]{\begingroup \@sanitize@url \@href}%
	\providecommand \@href[1]{\@@startlink{#1}\@@href}%
	\providecommand \@@href[1]{\endgroup#1\@@endlink}%
	\providecommand \@sanitize@url [0]{\catcode `\\12\catcode `\$12\catcode
		`\&12\catcode `\#12\catcode `\^12\catcode `\_12\catcode `\%12\relax}%
	\providecommand \@@startlink[1]{}%
	\providecommand \@@endlink[0]{}%
	\providecommand \url  [0]{\begingroup\@sanitize@url \@url }%
	\providecommand \@url [1]{\endgroup\@href {#1}{\urlprefix }}%
	\providecommand \urlprefix  [0]{URL }%
	\providecommand \Eprint [0]{\href }%
	\providecommand \doibase [0]{https://doi.org/}%
	\providecommand \selectlanguage [0]{\@gobble}%
	\providecommand \bibinfo  [0]{\@secondoftwo}%
	\providecommand \bibfield  [0]{\@secondoftwo}%
	\providecommand \translation [1]{[#1]}%
	\providecommand \BibitemOpen [0]{}%
	\providecommand \bibitemStop [0]{}%
	\providecommand \bibitemNoStop [0]{.\EOS\space}%
	\providecommand \EOS [0]{\spacefactor3000\relax}%
	\providecommand \BibitemShut  [1]{\csname bibitem#1\endcsname}%
	\let\auto@bib@innerbib\@empty
	%</preamble>
	\bibitem [{\citenamefont {Savary}\ and\ \citenamefont
		{Balents}(2016)}]{Savary016502}%
	\BibitemOpen
	\bibfield  {author} {\bibinfo {author} {\bibfnamefont {L.}~\bibnamefont
			{Savary}}\ and\ \bibinfo {author} {\bibfnamefont {L.}~\bibnamefont
			{Balents}},\ }\bibfield  {title} {\bibinfo {title} {{Quantum spin liquids: a
				review}},\ }\href {https://doi.org/10.1088/0034-4885/80/1/016502} {\bibfield
		{journal} {\bibinfo  {journal} {Rep. Prog. Phys.}\ }\textbf {\bibinfo
			{volume} {80}},\ \bibinfo {pages} {016502} (\bibinfo {year}
		{2016})}\BibitemShut {NoStop}%
	\bibitem [{\citenamefont {Bramwell}\ and\ \citenamefont
		{Gingras}(2001)}]{Bramwell1495}%
	\BibitemOpen
	\bibfield  {author} {\bibinfo {author} {\bibfnamefont {S.~T.}\ \bibnamefont
			{Bramwell}}\ and\ \bibinfo {author} {\bibfnamefont {M.~J.~P.}\ \bibnamefont
			{Gingras}},\ }\bibfield  {title} {\bibinfo {title} {{Spin Ice State in
				Frustrated Magnetic Pyrochlore Materials}},\ }\href
	{https://doi.org/10.1126/science.1064761} {\bibfield  {journal} {\bibinfo
			{journal} {Science}\ }\textbf {\bibinfo {volume} {294}},\ \bibinfo {pages}
		{1495} (\bibinfo {year} {2001})}\BibitemShut {NoStop}%
	\bibitem [{\citenamefont {Starykh}(2015)}]{Starykh052502}%
	\BibitemOpen
	\bibfield  {author} {\bibinfo {author} {\bibfnamefont {O.~A.}\ \bibnamefont
			{Starykh}},\ }\bibfield  {title} {\bibinfo {title} {{Unusual ordered phases
				of highly frustrated magnets: a review}},\ }\href
	{https://doi.org/10.1088/0034-4885/78/5/052502} {\bibfield  {journal}
		{\bibinfo  {journal} {Rep. Prog. Phys.}\ }\textbf {\bibinfo {volume} {78}},\
		\bibinfo {pages} {052502} (\bibinfo {year} {2015})}\BibitemShut {NoStop}%
	\bibitem [{\citenamefont {Rau}\ and\ \citenamefont {Gingras}(2019)}]{Rau357}%
	\BibitemOpen
	\bibfield  {author} {\bibinfo {author} {\bibfnamefont {J.~G.}\ \bibnamefont
			{Rau}}\ and\ \bibinfo {author} {\bibfnamefont {M.~J.}\ \bibnamefont
			{Gingras}},\ }\bibfield  {title} {\bibinfo {title} {{Frustrated Quantum
				Rare-Earth Pyrochlores}},\ }\href
	{https://doi.org/https://doi.org/10.1146/annurev-conmatphys-022317-110520}
	{\bibfield  {journal} {\bibinfo  {journal} {Annu. Rev. Condens. Matter
				Phys.}\ }\textbf {\bibinfo {volume} {10}},\ \bibinfo {pages} {357} (\bibinfo
		{year} {2019})}\BibitemShut {NoStop}%
	\bibitem [{\citenamefont {Graham}\ \emph {et~al.}(2023)\citenamefont {Graham},
		\citenamefont {Qureshi}, \citenamefont {Ritter}, \citenamefont {Manuel},
		\citenamefont {Wildes},\ and\ \citenamefont {Clark}}]{Graham166703}%
	\BibitemOpen
	\bibfield  {author} {\bibinfo {author} {\bibfnamefont {J.~N.}\ \bibnamefont
			{Graham}}, \bibinfo {author} {\bibfnamefont {N.}~\bibnamefont {Qureshi}},
		\bibinfo {author} {\bibfnamefont {C.}~\bibnamefont {Ritter}}, \bibinfo
		{author} {\bibfnamefont {P.}~\bibnamefont {Manuel}}, \bibinfo {author}
		{\bibfnamefont {A.~R.}\ \bibnamefont {Wildes}},\ and\ \bibinfo {author}
		{\bibfnamefont {L.}~\bibnamefont {Clark}},\ }\bibfield  {title} {\bibinfo
		{title} {{Experimental Evidence for the Spiral Spin Liquid in
				${\mathrm{LiYbO}}_{2}$}},\ }\href
	{https://doi.org/10.1103/PhysRevLett.130.166703} {\bibfield  {journal}
		{\bibinfo  {journal} {Phys. Rev. Lett.}\ }\textbf {\bibinfo {volume} {130}},\
		\bibinfo {pages} {166703} (\bibinfo {year} {2023})}\BibitemShut {NoStop}%
	\bibitem [{\citenamefont {Bordelon}\ \emph {et~al.}(2019)\citenamefont
		{Bordelon}, \citenamefont {Kenney}, \citenamefont {Liu}, \citenamefont
		{Hogan}, \citenamefont {Posthuma}, \citenamefont {Kavand}, \citenamefont
		{Lyu}, \citenamefont {Sherwin}, \citenamefont {Butch}, \citenamefont {Brown},
		\citenamefont {Graf}, \citenamefont {Balents},\ and\ \citenamefont
		{Wilson}}]{Bordelon1058}%
	\BibitemOpen
	\bibfield  {author} {\bibinfo {author} {\bibfnamefont {M.~M.}\ \bibnamefont
			{Bordelon}}, \bibinfo {author} {\bibfnamefont {E.}~\bibnamefont {Kenney}},
		\bibinfo {author} {\bibfnamefont {C.}~\bibnamefont {Liu}}, \bibinfo {author}
		{\bibfnamefont {T.}~\bibnamefont {Hogan}}, \bibinfo {author} {\bibfnamefont
			{L.}~\bibnamefont {Posthuma}}, \bibinfo {author} {\bibfnamefont
			{M.}~\bibnamefont {Kavand}}, \bibinfo {author} {\bibfnamefont
			{Y.}~\bibnamefont {Lyu}}, \bibinfo {author} {\bibfnamefont {M.}~\bibnamefont
			{Sherwin}}, \bibinfo {author} {\bibfnamefont {N.~P.}\ \bibnamefont {Butch}},
		\bibinfo {author} {\bibfnamefont {C.}~\bibnamefont {Brown}}, \bibinfo
		{author} {\bibfnamefont {M.~J.}\ \bibnamefont {Graf}}, \bibinfo {author}
		{\bibfnamefont {L.}~\bibnamefont {Balents}},\ and\ \bibinfo {author}
		{\bibfnamefont {S.~D.}\ \bibnamefont {Wilson}},\ }\bibfield  {title}
	{\bibinfo {title} {{Field-tunable quantum disordered ground state in the
				triangular-lattice antiferromagnet NaYbO$_2$}},\ }\href
	{https://doi.org/10.1038/s41567-019-0594-5} {\bibfield  {journal} {\bibinfo
			{journal} {Nat. Phys.}\ }\textbf {\bibinfo {volume} {15}},\ \bibinfo {pages}
		{1058} (\bibinfo {year} {2019})}\BibitemShut {NoStop}%
	\bibitem [{\citenamefont {Li}\ \emph {et~al.}(2017)\citenamefont {Li},
		\citenamefont {Adroja}, \citenamefont {Bewley}, \citenamefont {Voneshen},
		\citenamefont {Tsirlin}, \citenamefont {Gegenwart},\ and\ \citenamefont
		{Zhang}}]{Li107202}%
	\BibitemOpen
	\bibfield  {author} {\bibinfo {author} {\bibfnamefont {Y.}~\bibnamefont
			{Li}}, \bibinfo {author} {\bibfnamefont {D.}~\bibnamefont {Adroja}}, \bibinfo
		{author} {\bibfnamefont {R.~I.}\ \bibnamefont {Bewley}}, \bibinfo {author}
		{\bibfnamefont {D.}~\bibnamefont {Voneshen}}, \bibinfo {author}
		{\bibfnamefont {A.~A.}\ \bibnamefont {Tsirlin}}, \bibinfo {author}
		{\bibfnamefont {P.}~\bibnamefont {Gegenwart}},\ and\ \bibinfo {author}
		{\bibfnamefont {Q.}~\bibnamefont {Zhang}},\ }\bibfield  {title} {\bibinfo
		{title} {{Crystalline Electric-Field Randomness in the Triangular Lattice
				Spin-Liquid ${\mathrm{YbMgGaO}}_{4}$}},\ }\href
	{https://doi.org/10.1103/PhysRevLett.118.107202} {\bibfield  {journal}
		{\bibinfo  {journal} {Phys. Rev. Lett.}\ }\textbf {\bibinfo {volume} {118}},\
		\bibinfo {pages} {107202} (\bibinfo {year} {2017})}\BibitemShut {NoStop}%
	\bibitem [{\citenamefont {Sibille}\ \emph {et~al.}(2018)\citenamefont
		{Sibille}, \citenamefont {Gauthier}, \citenamefont {Yan}, \citenamefont
		{Ciomaga~Hatnean}, \citenamefont {Ollivier}, \citenamefont {Winn},
		\citenamefont {Filges}, \citenamefont {Balakrishnan}, \citenamefont
		{Kenzelmann}, \citenamefont {Shannon},\ and\ \citenamefont
		{Fennell}}]{Sibille711}%
	\BibitemOpen
	\bibfield  {author} {\bibinfo {author} {\bibfnamefont {R.}~\bibnamefont
			{Sibille}}, \bibinfo {author} {\bibfnamefont {N.}~\bibnamefont {Gauthier}},
		\bibinfo {author} {\bibfnamefont {H.}~\bibnamefont {Yan}}, \bibinfo {author}
		{\bibfnamefont {M.}~\bibnamefont {Ciomaga~Hatnean}}, \bibinfo {author}
		{\bibfnamefont {J.}~\bibnamefont {Ollivier}}, \bibinfo {author}
		{\bibfnamefont {B.}~\bibnamefont {Winn}}, \bibinfo {author} {\bibfnamefont
			{U.}~\bibnamefont {Filges}}, \bibinfo {author} {\bibfnamefont
			{G.}~\bibnamefont {Balakrishnan}}, \bibinfo {author} {\bibfnamefont
			{M.}~\bibnamefont {Kenzelmann}}, \bibinfo {author} {\bibfnamefont
			{N.}~\bibnamefont {Shannon}},\ and\ \bibinfo {author} {\bibfnamefont
			{T.}~\bibnamefont {Fennell}},\ }\bibfield  {title} {\bibinfo {title}
		{{Experimental signatures of emergent quantum electrodynamics in
				Pr$_2$Hf$_2$O$_7$}},\ }\href {https://doi.org/10.1038/s41567-018-0116-x}
	{\bibfield  {journal} {\bibinfo  {journal} {Nat. Phys.}\ }\textbf {\bibinfo
			{volume} {14}},\ \bibinfo {pages} {711} (\bibinfo {year} {2018})}\BibitemShut
	{NoStop}%
	\bibitem [{\citenamefont {Guchhait}\ \emph {et~al.}(2024)\citenamefont
		{Guchhait}, \citenamefont {Painganoor}, \citenamefont {Islam}, \citenamefont
		{Sichelschmidt}, \citenamefont {Le}, \citenamefont {Aouane}, \citenamefont
		{Christensen},\ and\ \citenamefont {Nath}}]{Guchhait144434}%
	\BibitemOpen
	\bibfield  {author} {\bibinfo {author} {\bibfnamefont {S.}~\bibnamefont
			{Guchhait}}, \bibinfo {author} {\bibfnamefont {A.}~\bibnamefont
			{Painganoor}}, \bibinfo {author} {\bibfnamefont {S.~S.}\ \bibnamefont
			{Islam}}, \bibinfo {author} {\bibfnamefont {J.}~\bibnamefont
			{Sichelschmidt}}, \bibinfo {author} {\bibfnamefont {M.~D.}\ \bibnamefont
			{Le}}, \bibinfo {author} {\bibfnamefont {M.}~\bibnamefont {Aouane}}, \bibinfo
		{author} {\bibfnamefont {N.~B.}\ \bibnamefont {Christensen}},\ and\ \bibinfo
		{author} {\bibfnamefont {R.}~\bibnamefont {Nath}},\ }\bibfield  {title}
	{\bibinfo {title} {{Magnetic and crystal electric field studies of the rare
				earth based square lattice antiferromagnet ${\mathrm{NdKNaNbO}}_{5}$}},\
	}\href {https://doi.org/10.1103/PhysRevB.110.144434} {\bibfield  {journal}
		{\bibinfo  {journal} {Phys. Rev. B}\ }\textbf {\bibinfo {volume} {110}},\
		\bibinfo {pages} {144434} (\bibinfo {year} {2024})}\BibitemShut {NoStop}%
	\bibitem [{\citenamefont {Zhang}\ \emph {et~al.}(2025)\citenamefont {Zhang},
		\citenamefont {Shu}, \citenamefont {Xie}, \citenamefont {Zhuo}, \citenamefont
		{Cai}, \citenamefont {Balz}, \citenamefont {Ji}, \citenamefont {Jin},
		\citenamefont {Ma},\ and\ \citenamefont {Zhang}}]{Zhang256503}%
	\BibitemOpen
	\bibfield  {author} {\bibinfo {author} {\bibfnamefont {Z.}~\bibnamefont
			{Zhang}}, \bibinfo {author} {\bibfnamefont {M.}~\bibnamefont {Shu}}, \bibinfo
		{author} {\bibfnamefont {M.}~\bibnamefont {Xie}}, \bibinfo {author}
		{\bibfnamefont {W.}~\bibnamefont {Zhuo}}, \bibinfo {author} {\bibfnamefont
			{Y.}~\bibnamefont {Cai}}, \bibinfo {author} {\bibfnamefont {C.}~\bibnamefont
			{Balz}}, \bibinfo {author} {\bibfnamefont {J.}~\bibnamefont {Ji}}, \bibinfo
		{author} {\bibfnamefont {F.}~\bibnamefont {Jin}}, \bibinfo {author}
		{\bibfnamefont {J.}~\bibnamefont {Ma}},\ and\ \bibinfo {author}
		{\bibfnamefont {Q.}~\bibnamefont {Zhang}},\ }\bibfield  {title} {\bibinfo
		{title} {{Emergent Dispersive Multipolar Excitations in
				${\mathrm{NaErSe}}_{2}$}},\ }\href {https://doi.org/10.1103/7k9l-j1kh}
	{\bibfield  {journal} {\bibinfo  {journal} {Phys. Rev. Lett.}\ }\textbf
		{\bibinfo {volume} {135}},\ \bibinfo {pages} {256503} (\bibinfo {year}
		{2025})}\BibitemShut {NoStop}%
	\bibitem [{\citenamefont {Ranjith}\ \emph {et~al.}(2019)\citenamefont
		{Ranjith}, \citenamefont {Dmytriieva}, \citenamefont {Khim}, \citenamefont
		{Sichelschmidt}, \citenamefont {Luther}, \citenamefont {Ehlers},
		\citenamefont {Yasuoka}, \citenamefont {Wosnitza}, \citenamefont {Tsirlin},
		\citenamefont {K\"uhne},\ and\ \citenamefont {Baenitz}}]{Ranjith180401}%
	\BibitemOpen
	\bibfield  {author} {\bibinfo {author} {\bibfnamefont {K.~M.}\ \bibnamefont
			{Ranjith}}, \bibinfo {author} {\bibfnamefont {D.}~\bibnamefont {Dmytriieva}},
		\bibinfo {author} {\bibfnamefont {S.}~\bibnamefont {Khim}}, \bibinfo {author}
		{\bibfnamefont {J.}~\bibnamefont {Sichelschmidt}}, \bibinfo {author}
		{\bibfnamefont {S.}~\bibnamefont {Luther}}, \bibinfo {author} {\bibfnamefont
			{D.}~\bibnamefont {Ehlers}}, \bibinfo {author} {\bibfnamefont
			{H.}~\bibnamefont {Yasuoka}}, \bibinfo {author} {\bibfnamefont
			{J.}~\bibnamefont {Wosnitza}}, \bibinfo {author} {\bibfnamefont {A.~A.}\
			\bibnamefont {Tsirlin}}, \bibinfo {author} {\bibfnamefont {H.}~\bibnamefont
			{K\"uhne}},\ and\ \bibinfo {author} {\bibfnamefont {M.}~\bibnamefont
			{Baenitz}},\ }\bibfield  {title} {\bibinfo {title} {{Field-induced
				instability of the quantum spin liquid ground state in the
				${J}_{\mathrm{eff}}=\frac{1}{2}$ triangular-lattice compound
				${\mathrm{NaYbO}}_{2}$}},\ }\href
	{https://doi.org/10.1103/PhysRevB.99.180401} {\bibfield  {journal} {\bibinfo
			{journal} {Phys. Rev. B}\ }\textbf {\bibinfo {volume} {99}},\ \bibinfo
		{pages} {180401(R)} (\bibinfo {year} {2019})}\BibitemShut {NoStop}%
	\bibitem [{\citenamefont {Yamamoto}\ \emph {et~al.}(2023)\citenamefont
		{Yamamoto}, \citenamefont {Le}, \citenamefont {Adroja}, \citenamefont
		{Shimura}, \citenamefont {Takabatake},\ and\ \citenamefont
		{Onimaru}}]{Yamamoto075114}%
	\BibitemOpen
	\bibfield  {author} {\bibinfo {author} {\bibfnamefont {R.}~\bibnamefont
			{Yamamoto}}, \bibinfo {author} {\bibfnamefont {M.~D.}\ \bibnamefont {Le}},
		\bibinfo {author} {\bibfnamefont {D.~T.}\ \bibnamefont {Adroja}}, \bibinfo
		{author} {\bibfnamefont {Y.}~\bibnamefont {Shimura}}, \bibinfo {author}
		{\bibfnamefont {T.}~\bibnamefont {Takabatake}},\ and\ \bibinfo {author}
		{\bibfnamefont {T.}~\bibnamefont {Onimaru}},\ }\bibfield  {title} {\bibinfo
		{title} {{Inelastic neutron scattering study of crystalline electric field
				excitations in the caged compounds $\mathrm{Nd}{T}_{2}{\mathrm{Zn}}_{20}
				(T=\mathrm{Co}, \mathrm{Rh}, \text{and} \mathrm{Ir})$}},\ }\href
	{https://doi.org/10.1103/PhysRevB.107.075114} {\bibfield  {journal} {\bibinfo
			{journal} {Phys. Rev. B}\ }\textbf {\bibinfo {volume} {107}},\ \bibinfo
		{pages} {075114} (\bibinfo {year} {2023})}\BibitemShut {NoStop}%
	\bibitem [{\citenamefont {Rau}\ and\ \citenamefont
		{Gingras}(2015)}]{Rau144417}%
	\BibitemOpen
	\bibfield  {author} {\bibinfo {author} {\bibfnamefont {J.~G.}\ \bibnamefont
			{Rau}}\ and\ \bibinfo {author} {\bibfnamefont {M.~J.~P.}\ \bibnamefont
			{Gingras}},\ }\bibfield  {title} {\bibinfo {title} {Magnitude of quantum
			effects in classical spin ices},\ }\href
	{https://doi.org/10.1103/PhysRevB.92.144417} {\bibfield  {journal} {\bibinfo
			{journal} {Phys. Rev. B}\ }\textbf {\bibinfo {volume} {92}},\ \bibinfo
		{pages} {144417} (\bibinfo {year} {2015})}\BibitemShut {NoStop}%
	\bibitem [{\citenamefont {Tomasello}\ \emph {et~al.}(2015)\citenamefont
		{Tomasello}, \citenamefont {Castelnovo}, \citenamefont {Moessner},\ and\
		\citenamefont {Quintanilla}}]{Tomasello155120}%
	\BibitemOpen
	\bibfield  {author} {\bibinfo {author} {\bibfnamefont {B.}~\bibnamefont
			{Tomasello}}, \bibinfo {author} {\bibfnamefont {C.}~\bibnamefont
			{Castelnovo}}, \bibinfo {author} {\bibfnamefont {R.}~\bibnamefont
			{Moessner}},\ and\ \bibinfo {author} {\bibfnamefont {J.}~\bibnamefont
			{Quintanilla}},\ }\bibfield  {title} {\bibinfo {title} {{Single-ion
				anisotropy and magnetic field response in the spin-ice materials
				${\mathrm{Ho}}_{2}{\mathrm{Ti}}_{2}{\mathrm{O}}_{7}$ and
				${\mathrm{Dy}}_{2}{\mathrm{Ti}}_{2}{\mathrm{O}}_{7}$}},\ }\href
	{https://doi.org/10.1103/PhysRevB.92.155120} {\bibfield  {journal} {\bibinfo
			{journal} {Phys. Rev. B}\ }\textbf {\bibinfo {volume} {92}},\ \bibinfo
		{pages} {155120} (\bibinfo {year} {2015})}\BibitemShut {NoStop}%
	\bibitem [{\citenamefont {Gao}\ \emph {et~al.}(2020)\citenamefont {Gao},
		\citenamefont {Xiao}, \citenamefont {Kamazawa}, \citenamefont {Ikeuchi},
		\citenamefont {Biner}, \citenamefont {Kr\"amer}, \citenamefont {R\"uegg},\
		and\ \citenamefont {Arima}}]{Gao024424}%
	\BibitemOpen
	\bibfield  {author} {\bibinfo {author} {\bibfnamefont {S.}~\bibnamefont
			{Gao}}, \bibinfo {author} {\bibfnamefont {F.}~\bibnamefont {Xiao}}, \bibinfo
		{author} {\bibfnamefont {K.}~\bibnamefont {Kamazawa}}, \bibinfo {author}
		{\bibfnamefont {K.}~\bibnamefont {Ikeuchi}}, \bibinfo {author} {\bibfnamefont
			{D.}~\bibnamefont {Biner}}, \bibinfo {author} {\bibfnamefont {K.~W.}\
			\bibnamefont {Kr\"amer}}, \bibinfo {author} {\bibfnamefont {C.}~\bibnamefont
			{R\"uegg}},\ and\ \bibinfo {author} {\bibfnamefont {T.-h.}\ \bibnamefont
			{Arima}},\ }\bibfield  {title} {\bibinfo {title} {{Crystal electric field
				excitations in the quantum spin liquid candidate ${\mathrm{NaErS}}_{2}$}},\
	}\href {https://doi.org/10.1103/PhysRevB.102.024424} {\bibfield  {journal}
		{\bibinfo  {journal} {Phys. Rev. B}\ }\textbf {\bibinfo {volume} {102}},\
		\bibinfo {pages} {024424} (\bibinfo {year} {2020})}\BibitemShut {NoStop}%
	\bibitem [{\citenamefont {Scheie}\ \emph {et~al.}(2020)\citenamefont {Scheie},
		\citenamefont {Garlea}, \citenamefont {Sanjeewa}, \citenamefont {Xing},\ and\
		\citenamefont {Sefat}}]{Scheie144432}%
	\BibitemOpen
	\bibfield  {author} {\bibinfo {author} {\bibfnamefont {A.}~\bibnamefont
			{Scheie}}, \bibinfo {author} {\bibfnamefont {V.~O.}\ \bibnamefont {Garlea}},
		\bibinfo {author} {\bibfnamefont {L.~D.}\ \bibnamefont {Sanjeewa}}, \bibinfo
		{author} {\bibfnamefont {J.}~\bibnamefont {Xing}},\ and\ \bibinfo {author}
		{\bibfnamefont {A.~S.}\ \bibnamefont {Sefat}},\ }\bibfield  {title} {\bibinfo
		{title} {{Crystal-field Hamiltonian and anisotropy in ${\mathrm{KErSe}}_{2}$
				and ${\mathrm{CsErSe}}_{2}$}},\ }\href
	{https://doi.org/10.1103/PhysRevB.101.144432} {\bibfield  {journal} {\bibinfo
			{journal} {Phys. Rev. B}\ }\textbf {\bibinfo {volume} {101}},\ \bibinfo
		{pages} {144432} (\bibinfo {year} {2020})}\BibitemShut {NoStop}%
	\bibitem [{\citenamefont {Bag}\ \emph {et~al.}(2024)\citenamefont {Bag},
		\citenamefont {Xu}, \citenamefont {Sherman}, \citenamefont {Yadav},
		\citenamefont {Kolesnikov}, \citenamefont {Podlesnyak}, \citenamefont {Choi},
		\citenamefont {da~Silva}, \citenamefont {Moore},\ and\ \citenamefont
		{Haravifard}}]{Bag266703}%
	\BibitemOpen
	\bibfield  {author} {\bibinfo {author} {\bibfnamefont {R.}~\bibnamefont
			{Bag}}, \bibinfo {author} {\bibfnamefont {S.}~\bibnamefont {Xu}}, \bibinfo
		{author} {\bibfnamefont {N.~E.}\ \bibnamefont {Sherman}}, \bibinfo {author}
		{\bibfnamefont {L.}~\bibnamefont {Yadav}}, \bibinfo {author} {\bibfnamefont
			{A.~I.}\ \bibnamefont {Kolesnikov}}, \bibinfo {author} {\bibfnamefont
			{A.~A.}\ \bibnamefont {Podlesnyak}}, \bibinfo {author} {\bibfnamefont
			{E.~S.}\ \bibnamefont {Choi}}, \bibinfo {author} {\bibfnamefont
			{I.}~\bibnamefont {da~Silva}}, \bibinfo {author} {\bibfnamefont {J.~E.}\
			\bibnamefont {Moore}},\ and\ \bibinfo {author} {\bibfnamefont
			{S.}~\bibnamefont {Haravifard}},\ }\bibfield  {title} {\bibinfo {title}
		{{Evidence of Dirac Quantum Spin Liquid in
				${\mathrm{YbZn}}_{2}{\mathrm{GaO}}_{5}$}},\ }\href
	{https://doi.org/10.1103/PhysRevLett.133.266703} {\bibfield  {journal}
		{\bibinfo  {journal} {Phys. Rev. Lett.}\ }\textbf {\bibinfo {volume} {133}},\
		\bibinfo {pages} {266703} (\bibinfo {year} {2024})}\BibitemShut {NoStop}%
	\bibitem [{\citenamefont {Clark}\ \emph {et~al.}(2019)\citenamefont {Clark},
		\citenamefont {Sala}, \citenamefont {Maharaj}, \citenamefont {Stone},
		\citenamefont {Knight}, \citenamefont {Telling}, \citenamefont {Wang},
		\citenamefont {Xu}, \citenamefont {Kim}, \citenamefont {Li}, \citenamefont
		{Cheong},\ and\ \citenamefont {Gaulin}}]{Clark262}%
	\BibitemOpen
	\bibfield  {author} {\bibinfo {author} {\bibfnamefont {L.}~\bibnamefont
			{Clark}}, \bibinfo {author} {\bibfnamefont {G.}~\bibnamefont {Sala}},
		\bibinfo {author} {\bibfnamefont {D.~D.}\ \bibnamefont {Maharaj}}, \bibinfo
		{author} {\bibfnamefont {M.~B.}\ \bibnamefont {Stone}}, \bibinfo {author}
		{\bibfnamefont {K.~S.}\ \bibnamefont {Knight}}, \bibinfo {author}
		{\bibfnamefont {M.~T.~F.}\ \bibnamefont {Telling}}, \bibinfo {author}
		{\bibfnamefont {X.}~\bibnamefont {Wang}}, \bibinfo {author} {\bibfnamefont
			{X.}~\bibnamefont {Xu}}, \bibinfo {author} {\bibfnamefont {J.}~\bibnamefont
			{Kim}}, \bibinfo {author} {\bibfnamefont {Y.}~\bibnamefont {Li}}, \bibinfo
		{author} {\bibfnamefont {S.-W.}\ \bibnamefont {Cheong}},\ and\ \bibinfo
		{author} {\bibfnamefont {B.~D.}\ \bibnamefont {Gaulin}},\ }\bibfield  {title}
	{\bibinfo {title} {{Two-dimensional spin liquid behaviour in the
				triangular-honeycomb antiferromagnet TbInO$_3$}},\ }\href
	{https://doi.org/10.1038/s41567-018-0407-2} {\bibfield  {journal} {\bibinfo
			{journal} {Nat. Phys.}\ }\textbf {\bibinfo {volume} {15}},\ \bibinfo {pages}
		{262} (\bibinfo {year} {2019})}\BibitemShut {NoStop}%
	\bibitem [{\citenamefont {Somesh}\ \emph {et~al.}(2023)\citenamefont {Somesh},
		\citenamefont {Islam}, \citenamefont {Mohanty}, \citenamefont {Simutis},
		\citenamefont {Guguchia}, \citenamefont {Wang}, \citenamefont
		{Sichelschmidt}, \citenamefont {Baenitz},\ and\ \citenamefont
		{Nath}}]{Somesh064421}%
	\BibitemOpen
	\bibfield  {author} {\bibinfo {author} {\bibfnamefont {K.}~\bibnamefont
			{Somesh}}, \bibinfo {author} {\bibfnamefont {S.~S.}\ \bibnamefont {Islam}},
		\bibinfo {author} {\bibfnamefont {S.}~\bibnamefont {Mohanty}}, \bibinfo
		{author} {\bibfnamefont {G.}~\bibnamefont {Simutis}}, \bibinfo {author}
		{\bibfnamefont {Z.}~\bibnamefont {Guguchia}}, \bibinfo {author}
		{\bibfnamefont {C.}~\bibnamefont {Wang}}, \bibinfo {author} {\bibfnamefont
			{J.}~\bibnamefont {Sichelschmidt}}, \bibinfo {author} {\bibfnamefont
			{M.}~\bibnamefont {Baenitz}},\ and\ \bibinfo {author} {\bibfnamefont
			{R.}~\bibnamefont {Nath}},\ }\bibfield  {title} {\bibinfo {title} {{Absence
				of magnetic order and emergence of unconventional fluctuations in the
				${J}_{\mathrm{eff}}=\frac{1}{2}$ triangular-lattice antiferromagnet
				${\mathrm{YbBO}}_{3}$}},\ }\href
	{https://doi.org/10.1103/PhysRevB.107.064421} {\bibfield  {journal} {\bibinfo
			{journal} {Phys. Rev. B}\ }\textbf {\bibinfo {volume} {107}},\ \bibinfo
		{pages} {064421} (\bibinfo {year} {2023})}\BibitemShut {NoStop}%
	\bibitem [{\citenamefont {Mohanty}\ \emph {et~al.}(2026)\citenamefont
		{Mohanty}, \citenamefont {Guchhait}, \citenamefont {Islam}, \citenamefont
		{Patra}, \citenamefont {Saravanan}, \citenamefont {Krieger}, \citenamefont
		{Hicken}, \citenamefont {Luetkens}, \citenamefont {Adroja}, \citenamefont
		{Nilsen}, \citenamefont {Le},\ and\ \citenamefont {Nath}}]{Mohanty214452}%
	\BibitemOpen
	\bibfield  {author} {\bibinfo {author} {\bibfnamefont {S.}~\bibnamefont
			{Mohanty}}, \bibinfo {author} {\bibfnamefont {S.}~\bibnamefont {Guchhait}},
		\bibinfo {author} {\bibfnamefont {S.~S.}\ \bibnamefont {Islam}}, \bibinfo
		{author} {\bibfnamefont {S.~P.}\ \bibnamefont {Patra}}, \bibinfo {author}
		{\bibfnamefont {M.~P.}\ \bibnamefont {Saravanan}}, \bibinfo {author}
		{\bibfnamefont {J.~A.}\ \bibnamefont {Krieger}}, \bibinfo {author}
		{\bibfnamefont {T.~J.}\ \bibnamefont {Hicken}}, \bibinfo {author}
		{\bibfnamefont {H.}~\bibnamefont {Luetkens}}, \bibinfo {author}
		{\bibfnamefont {D.~T.}\ \bibnamefont {Adroja}}, \bibinfo {author}
		{\bibfnamefont {G.~J.}\ \bibnamefont {Nilsen}}, \bibinfo {author}
		{\bibfnamefont {M.~D.}\ \bibnamefont {Le}},\ and\ \bibinfo {author}
		{\bibfnamefont {R.}~\bibnamefont {Nath}},\ }\bibfield  {title} {\bibinfo
		{title} {{Crystal electric field excitations and spin dynamics in the
				spin-orbit coupled distorted honeycomb magnet ${\mathrm{BiErGeO}}_{5}$}},\
	}\href {https://doi.org/10.1103/z66x-362w} {\bibfield  {journal} {\bibinfo
			{journal} {Phys. Rev. B}\ }\textbf {\bibinfo {volume} {113}},\ \bibinfo
		{pages} {214452} (\bibinfo {year} {2026})}\BibitemShut {NoStop}%
	\bibitem [{\citenamefont {Gao}\ \emph {et~al.}(2019)\citenamefont {Gao},
		\citenamefont {Chen}, \citenamefont {Tam}, \citenamefont {Huang},
		\citenamefont {Sasmal}, \citenamefont {Adroja}, \citenamefont {Ye},
		\citenamefont {Cao}, \citenamefont {Sala}, \citenamefont {Stone},
		\citenamefont {Baines}, \citenamefont {Verezhak}, \citenamefont {Hu},
		\citenamefont {Chung}, \citenamefont {Xu}, \citenamefont {Cheong},
		\citenamefont {Nallaiyan}, \citenamefont {Spagna}, \citenamefont {Maple},
		\citenamefont {Nevidomskyy}, \citenamefont {Morosan}, \citenamefont {Chen},\
		and\ \citenamefont {Dai}}]{Gao1052}%
	\BibitemOpen
	\bibfield  {author} {\bibinfo {author} {\bibfnamefont {B.}~\bibnamefont
			{Gao}}, \bibinfo {author} {\bibfnamefont {T.}~\bibnamefont {Chen}}, \bibinfo
		{author} {\bibfnamefont {D.~W.}\ \bibnamefont {Tam}}, \bibinfo {author}
		{\bibfnamefont {C.-L.}\ \bibnamefont {Huang}}, \bibinfo {author}
		{\bibfnamefont {K.}~\bibnamefont {Sasmal}}, \bibinfo {author} {\bibfnamefont
			{D.~T.}\ \bibnamefont {Adroja}}, \bibinfo {author} {\bibfnamefont
			{F.}~\bibnamefont {Ye}}, \bibinfo {author} {\bibfnamefont {H.}~\bibnamefont
			{Cao}}, \bibinfo {author} {\bibfnamefont {G.}~\bibnamefont {Sala}}, \bibinfo
		{author} {\bibfnamefont {M.~B.}\ \bibnamefont {Stone}}, \bibinfo {author}
		{\bibfnamefont {C.}~\bibnamefont {Baines}}, \bibinfo {author} {\bibfnamefont
			{J.~A.~T.}\ \bibnamefont {Verezhak}}, \bibinfo {author} {\bibfnamefont
			{H.}~\bibnamefont {Hu}}, \bibinfo {author} {\bibfnamefont {J.-H.}\
			\bibnamefont {Chung}}, \bibinfo {author} {\bibfnamefont {X.}~\bibnamefont
			{Xu}}, \bibinfo {author} {\bibfnamefont {S.-W.}\ \bibnamefont {Cheong}},
		\bibinfo {author} {\bibfnamefont {M.}~\bibnamefont {Nallaiyan}}, \bibinfo
		{author} {\bibfnamefont {S.}~\bibnamefont {Spagna}}, \bibinfo {author}
		{\bibfnamefont {M.~B.}\ \bibnamefont {Maple}}, \bibinfo {author}
		{\bibfnamefont {A.~H.}\ \bibnamefont {Nevidomskyy}}, \bibinfo {author}
		{\bibfnamefont {E.}~\bibnamefont {Morosan}}, \bibinfo {author} {\bibfnamefont
			{G.}~\bibnamefont {Chen}},\ and\ \bibinfo {author} {\bibfnamefont
			{P.}~\bibnamefont {Dai}},\ }\bibfield  {title} {\bibinfo {title}
		{{Experimental signatures of a three-dimensional quantum spin liquid in
				effective spin-1/2 Ce$_2$Zr$_2$O$_7$ pyrochlore}},\ }\href
	{https://doi.org/10.1038/s41567-019-0577-6} {\bibfield  {journal} {\bibinfo
			{journal} {Nat. Phys.}\ }\textbf {\bibinfo {volume} {15}},\ \bibinfo {pages}
		{1052} (\bibinfo {year} {2019})}\BibitemShut {NoStop}%
	\bibitem [{\citenamefont {Kermarrec}\ \emph {et~al.}(2017)\citenamefont
		{Kermarrec}, \citenamefont {Gaudet}, \citenamefont {Fritsch}, \citenamefont
		{Khasanov}, \citenamefont {Guguchia}, \citenamefont {Ritter}, \citenamefont
		{Ross}, \citenamefont {Dabkowska},\ and\ \citenamefont
		{Gaulin}}]{Kermarrec14810}%
	\BibitemOpen
	\bibfield  {author} {\bibinfo {author} {\bibfnamefont {E.}~\bibnamefont
			{Kermarrec}}, \bibinfo {author} {\bibfnamefont {J.}~\bibnamefont {Gaudet}},
		\bibinfo {author} {\bibfnamefont {K.}~\bibnamefont {Fritsch}}, \bibinfo
		{author} {\bibfnamefont {R.}~\bibnamefont {Khasanov}}, \bibinfo {author}
		{\bibfnamefont {Z.}~\bibnamefont {Guguchia}}, \bibinfo {author}
		{\bibfnamefont {C.}~\bibnamefont {Ritter}}, \bibinfo {author} {\bibfnamefont
			{K.~A.}\ \bibnamefont {Ross}}, \bibinfo {author} {\bibfnamefont {H.~A.}\
			\bibnamefont {Dabkowska}},\ and\ \bibinfo {author} {\bibfnamefont {B.~D.}\
			\bibnamefont {Gaulin}},\ }\bibfield  {title} {\bibinfo {title} {{Ground state
				selection under pressure in the quantum pyrochlore magnet
				Yb$_2$Ti$_2$O$_7$}},\ }\href {https://doi.org/10.1038/ncomms14810} {\bibfield
		{journal} {\bibinfo  {journal} {Nat. Commun.}\ }\textbf {\bibinfo {volume}
			{8}},\ \bibinfo {pages} {14810} (\bibinfo {year} {2017})}\BibitemShut
	{NoStop}%
	\bibitem [{\citenamefont {Hallas}\ \emph {et~al.}(2018)\citenamefont {Hallas},
		\citenamefont {Gaudet},\ and\ \citenamefont {Gaulin}}]{Hallas105}%
	\BibitemOpen
	\bibfield  {author} {\bibinfo {author} {\bibfnamefont {A.~M.}\ \bibnamefont
			{Hallas}}, \bibinfo {author} {\bibfnamefont {J.}~\bibnamefont {Gaudet}},\
		and\ \bibinfo {author} {\bibfnamefont {B.~D.}\ \bibnamefont {Gaulin}},\
	}\bibfield  {title} {\bibinfo {title} {{Experimental Insights into
				Ground-State Selection of Quantum XY Pyrochlores}},\ }\href
	{https://doi.org/https://doi.org/10.1146/annurev-conmatphys-031016-025218}
	{\bibfield  {journal} {\bibinfo  {journal} {Annu. Rev. Condens. Matter
				Phys.}\ }\textbf {\bibinfo {volume} {9}},\ \bibinfo {pages} {105} (\bibinfo
		{year} {2018})}\BibitemShut {NoStop}%
	\bibitem [{\citenamefont {Gardner}\ \emph {et~al.}(2010)\citenamefont
		{Gardner}, \citenamefont {Gingras},\ and\ \citenamefont
		{Greedan}}]{Gardner53}%
	\BibitemOpen
	\bibfield  {author} {\bibinfo {author} {\bibfnamefont {J.~S.}\ \bibnamefont
			{Gardner}}, \bibinfo {author} {\bibfnamefont {M.~J.~P.}\ \bibnamefont
			{Gingras}},\ and\ \bibinfo {author} {\bibfnamefont {J.~E.}\ \bibnamefont
			{Greedan}},\ }\bibfield  {title} {\bibinfo {title} {{Magnetic pyrochlore
				oxides}},\ }\href {https://doi.org/10.1103/RevModPhys.82.53} {\bibfield
		{journal} {\bibinfo  {journal} {Rev. Mod. Phys.}\ }\textbf {\bibinfo {volume}
			{82}},\ \bibinfo {pages} {53} (\bibinfo {year} {2010})}\BibitemShut {NoStop}%
	\bibitem [{\citenamefont {Petrenko}\ \emph {et~al.}(1998)\citenamefont
		{Petrenko}, \citenamefont {Ritter}, \citenamefont {Yethiraj},\ and\
		\citenamefont {McK~Paul}}]{Petrenko4570}%
	\BibitemOpen
	\bibfield  {author} {\bibinfo {author} {\bibfnamefont {O.~A.}\ \bibnamefont
			{Petrenko}}, \bibinfo {author} {\bibfnamefont {C.}~\bibnamefont {Ritter}},
		\bibinfo {author} {\bibfnamefont {M.}~\bibnamefont {Yethiraj}},\ and\
		\bibinfo {author} {\bibfnamefont {D.}~\bibnamefont {McK~Paul}},\ }\bibfield
	{title} {\bibinfo {title} {{Investigation of the Low-Temperature Spin-Liquid
				Behavior of the Frustrated Magnet Gadolinium Gallium Garnet}},\ }\href
	{https://doi.org/10.1103/PhysRevLett.80.4570} {\bibfield  {journal} {\bibinfo
			{journal} {Phys. Rev. Lett.}\ }\textbf {\bibinfo {volume} {80}},\ \bibinfo
		{pages} {4570} (\bibinfo {year} {1998})}\BibitemShut {NoStop}%
	\bibitem [{\citenamefont {Paddison}\ \emph {et~al.}(2015)\citenamefont
		{Paddison}, \citenamefont {Jacobsen}, \citenamefont {Petrenko}, \citenamefont
		{Fernández-Díaz}, \citenamefont {Deen},\ and\ \citenamefont
		{Goodwin}}]{Joseph179}%
	\BibitemOpen
	\bibfield  {author} {\bibinfo {author} {\bibfnamefont {J.~A.~M.}\
			\bibnamefont {Paddison}}, \bibinfo {author} {\bibfnamefont {H.}~\bibnamefont
			{Jacobsen}}, \bibinfo {author} {\bibfnamefont {O.~A.}\ \bibnamefont
			{Petrenko}}, \bibinfo {author} {\bibfnamefont {M.~T.}\ \bibnamefont
			{Fernández-Díaz}}, \bibinfo {author} {\bibfnamefont {P.~P.}\ \bibnamefont
			{Deen}},\ and\ \bibinfo {author} {\bibfnamefont {A.~L.}\ \bibnamefont
			{Goodwin}},\ }\bibfield  {title} {\bibinfo {title} {{Hidden order in
				spin-liquid Gd$_3$Ga$_5$O$_{12}$}},\ }\href
	{https://doi.org/10.1126/science.aaa5326} {\bibfield  {journal} {\bibinfo
			{journal} {Science}\ }\textbf {\bibinfo {volume} {350}},\ \bibinfo {pages}
		{179} (\bibinfo {year} {2015})}\BibitemShut {NoStop}%
	\bibitem [{\citenamefont {Raymond}\ \emph {et~al.}(2024)\citenamefont
		{Raymond}, \citenamefont {Lhotel}, \citenamefont {Riordan}, \citenamefont
		{Ressouche}, \citenamefont {Beauvois}, \citenamefont {Marin},\ and\
		\citenamefont {Zhitomirsky}}]{Raymond236701}%
	\BibitemOpen
	\bibfield  {author} {\bibinfo {author} {\bibfnamefont {S.}~\bibnamefont
			{Raymond}}, \bibinfo {author} {\bibfnamefont {E.}~\bibnamefont {Lhotel}},
		\bibinfo {author} {\bibfnamefont {E.}~\bibnamefont {Riordan}}, \bibinfo
		{author} {\bibfnamefont {E.}~\bibnamefont {Ressouche}}, \bibinfo {author}
		{\bibfnamefont {K.}~\bibnamefont {Beauvois}}, \bibinfo {author}
		{\bibfnamefont {C.}~\bibnamefont {Marin}},\ and\ \bibinfo {author}
		{\bibfnamefont {M.~E.}\ \bibnamefont {Zhitomirsky}},\ }\bibfield  {title}
	{\bibinfo {title} {{Uncommon Magnetic Ordering in the Quantum Magnet
				${\mathrm{Yb}}_{3}{\mathrm{Ga}}_{5}{\mathrm{O}}_{12}$}},\ }\href
	{https://doi.org/10.1103/PhysRevLett.133.236701} {\bibfield  {journal}
		{\bibinfo  {journal} {Phys. Rev. Lett.}\ }\textbf {\bibinfo {volume} {133}},\
		\bibinfo {pages} {236701} (\bibinfo {year} {2024})}\BibitemShut {NoStop}%
	\bibitem [{\citenamefont {Xin}\ \emph {et~al.}(2026)\citenamefont {Xin},
		\citenamefont {Rutherford}, \citenamefont {Li}, \citenamefont {Feng},
		\citenamefont {Liu}, \citenamefont {Zhao}, \citenamefont {Wang},
		\citenamefont {Liang}, \citenamefont {Zhou}, \citenamefont {Li},
		\citenamefont {Xu}, \citenamefont {Xie}, \citenamefont {Choi}, \citenamefont
		{Zhao}, \citenamefont {Ma}, \citenamefont {Zhou},\ and\ \citenamefont
		{Sun}}]{Xin014436}%
	\BibitemOpen
	\bibfield  {author} {\bibinfo {author} {\bibfnamefont {Y.~F.}\ \bibnamefont
			{Xin}}, \bibinfo {author} {\bibfnamefont {A.}~\bibnamefont {Rutherford}},
		\bibinfo {author} {\bibfnamefont {N.}~\bibnamefont {Li}}, \bibinfo {author}
		{\bibfnamefont {M.~L.}\ \bibnamefont {Feng}}, \bibinfo {author}
		{\bibfnamefont {Y.~J.}\ \bibnamefont {Liu}}, \bibinfo {author} {\bibfnamefont
			{Z.~Y.}\ \bibnamefont {Zhao}}, \bibinfo {author} {\bibfnamefont {Y.~Y.}\
			\bibnamefont {Wang}}, \bibinfo {author} {\bibfnamefont {H.}~\bibnamefont
			{Liang}}, \bibinfo {author} {\bibfnamefont {Y.}~\bibnamefont {Zhou}},
		\bibinfo {author} {\bibfnamefont {Q.~J.}\ \bibnamefont {Li}}, \bibinfo
		{author} {\bibfnamefont {M.~Y.}\ \bibnamefont {Xu}}, \bibinfo {author}
		{\bibfnamefont {W.}~\bibnamefont {Xie}}, \bibinfo {author} {\bibfnamefont
			{E.~S.}\ \bibnamefont {Choi}}, \bibinfo {author} {\bibfnamefont
			{X.}~\bibnamefont {Zhao}}, \bibinfo {author} {\bibfnamefont {J.}~\bibnamefont
			{Ma}}, \bibinfo {author} {\bibfnamefont {H.~D.}\ \bibnamefont {Zhou}},\ and\
		\bibinfo {author} {\bibfnamefont {X.~F.}\ \bibnamefont {Sun}},\ }\bibfield
	{title} {\bibinfo {title} {{Thermodynamics and heat transport of a
				${\mathrm{Yb}}_{3}{\mathrm{Sc}}_{2}{\mathrm{Ga}}_{3}{\mathrm{O}}_{12}$ single
				crystal: A quantum spin liquid candidate}},\ }\href
	{https://doi.org/10.1103/styt-544l} {\bibfield  {journal} {\bibinfo
			{journal} {Phys. Rev. B}\ }\textbf {\bibinfo {volume} {113}},\ \bibinfo
		{pages} {014436} (\bibinfo {year} {2026})}\BibitemShut {NoStop}%
	\bibitem [{\citenamefont {Cai}\ \emph {et~al.}(2019)\citenamefont {Cai},
		\citenamefont {Wilson}, \citenamefont {Beare}, \citenamefont {Lygouras},
		\citenamefont {Thomas}, \citenamefont {Yahne}, \citenamefont {Ross},
		\citenamefont {Taddei}, \citenamefont {Sala}, \citenamefont {Dabkowska},
		\citenamefont {Aczel},\ and\ \citenamefont {Luke}}]{Cai184415}%
	\BibitemOpen
	\bibfield  {author} {\bibinfo {author} {\bibfnamefont {Y.}~\bibnamefont
			{Cai}}, \bibinfo {author} {\bibfnamefont {M.~N.}\ \bibnamefont {Wilson}},
		\bibinfo {author} {\bibfnamefont {J.}~\bibnamefont {Beare}}, \bibinfo
		{author} {\bibfnamefont {C.}~\bibnamefont {Lygouras}}, \bibinfo {author}
		{\bibfnamefont {G.}~\bibnamefont {Thomas}}, \bibinfo {author} {\bibfnamefont
			{D.~R.}\ \bibnamefont {Yahne}}, \bibinfo {author} {\bibfnamefont
			{K.}~\bibnamefont {Ross}}, \bibinfo {author} {\bibfnamefont {K.~M.}\
			\bibnamefont {Taddei}}, \bibinfo {author} {\bibfnamefont {G.}~\bibnamefont
			{Sala}}, \bibinfo {author} {\bibfnamefont {H.~A.}\ \bibnamefont {Dabkowska}},
		\bibinfo {author} {\bibfnamefont {A.~A.}\ \bibnamefont {Aczel}},\ and\
		\bibinfo {author} {\bibfnamefont {G.~M.}\ \bibnamefont {Luke}},\ }\bibfield
	{title} {\bibinfo {title} {{Crystal fields and magnetic structure of the
				Ising antiferromagnet
				${\mathrm{Er}}_{3}{\mathrm{Ga}}_{5}{\mathrm{O}}_{12}$}},\ }\href
	{https://doi.org/10.1103/PhysRevB.100.184415} {\bibfield  {journal} {\bibinfo
			{journal} {Phys. Rev. B}\ }\textbf {\bibinfo {volume} {100}},\ \bibinfo
		{pages} {184415} (\bibinfo {year} {2019})}\BibitemShut {NoStop}%
	\bibitem [{\citenamefont {Petit}\ \emph {et~al.}(2021)\citenamefont {Petit},
		\citenamefont {Damay}, \citenamefont {Berrod},\ and\ \citenamefont
		{Zanotti}}]{Petit013030}%
	\BibitemOpen
	\bibfield  {author} {\bibinfo {author} {\bibfnamefont {S.}~\bibnamefont
			{Petit}}, \bibinfo {author} {\bibfnamefont {F.}~\bibnamefont {Damay}},
		\bibinfo {author} {\bibfnamefont {Q.}~\bibnamefont {Berrod}},\ and\ \bibinfo
		{author} {\bibfnamefont {J.~M.}\ \bibnamefont {Zanotti}},\ }\bibfield
	{title} {\bibinfo {title} {{Spin and lattice dynamics in the two-singlet
				system ${\mathrm{Tb}}_{3}{\mathrm{Ga}}_{5}{\mathrm{O}}_{12}$}},\ }\href
	{https://doi.org/10.1103/PhysRevResearch.3.013030} {\bibfield  {journal}
		{\bibinfo  {journal} {Phys. Rev. Res.}\ }\textbf {\bibinfo {volume} {3}},\
		\bibinfo {pages} {013030} (\bibinfo {year} {2021})}\BibitemShut {NoStop}%
	\bibitem [{\citenamefont {Arh}\ \emph {et~al.}(2022)\citenamefont {Arh},
		\citenamefont {Sana}, \citenamefont {Pregelj}, \citenamefont {Khuntia},
		\citenamefont {Jagli{\v{c}}i{\'{c}}}, \citenamefont {Le}, \citenamefont
		{Biswas}, \citenamefont {Manuel}, \citenamefont {Mangin-Thro}, \citenamefont
		{Ozarowski},\ and\ \citenamefont {Zorko}}]{Arh2022}%
	\BibitemOpen
	\bibfield  {author} {\bibinfo {author} {\bibfnamefont {T.}~\bibnamefont
			{Arh}}, \bibinfo {author} {\bibfnamefont {B.}~\bibnamefont {Sana}}, \bibinfo
		{author} {\bibfnamefont {M.}~\bibnamefont {Pregelj}}, \bibinfo {author}
		{\bibfnamefont {P.}~\bibnamefont {Khuntia}}, \bibinfo {author} {\bibfnamefont
			{Z.}~\bibnamefont {Jagli{\v{c}}i{\'{c}}}}, \bibinfo {author} {\bibfnamefont
			{M.~D.}\ \bibnamefont {Le}}, \bibinfo {author} {\bibfnamefont {P.~K.}\
			\bibnamefont {Biswas}}, \bibinfo {author} {\bibfnamefont {P.}~\bibnamefont
			{Manuel}}, \bibinfo {author} {\bibfnamefont {L.}~\bibnamefont {Mangin-Thro}},
		\bibinfo {author} {\bibfnamefont {A.}~\bibnamefont {Ozarowski}},\ and\
		\bibinfo {author} {\bibfnamefont {A.}~\bibnamefont {Zorko}},\ }\bibfield
	{title} {\bibinfo {title} {{The Ising triangular-lattice antiferromagnet
				neodymium heptatantalate as a quantum spin liquid candidate}},\ }\href
	{https://doi.org/10.1038/s41563-021-01169-y} {\bibfield  {journal} {\bibinfo
			{journal} {Nat. Mater.}\ }\textbf {\bibinfo {volume} {21}},\ \bibinfo {pages}
		{416} (\bibinfo {year} {2022})}\BibitemShut {NoStop}%
	\bibitem [{\citenamefont {Xu}\ \emph {et~al.}(2015)\citenamefont {Xu},
		\citenamefont {Anand}, \citenamefont {Bera}, \citenamefont {Frontzek},
		\citenamefont {Abernathy}, \citenamefont {Casati}, \citenamefont
		{Siemensmeyer},\ and\ \citenamefont {Lake}}]{Xu224430}%
	\BibitemOpen
	\bibfield  {author} {\bibinfo {author} {\bibfnamefont {J.}~\bibnamefont
			{Xu}}, \bibinfo {author} {\bibfnamefont {V.~K.}\ \bibnamefont {Anand}},
		\bibinfo {author} {\bibfnamefont {A.~K.}\ \bibnamefont {Bera}}, \bibinfo
		{author} {\bibfnamefont {M.}~\bibnamefont {Frontzek}}, \bibinfo {author}
		{\bibfnamefont {D.~L.}\ \bibnamefont {Abernathy}}, \bibinfo {author}
		{\bibfnamefont {N.}~\bibnamefont {Casati}}, \bibinfo {author} {\bibfnamefont
			{K.}~\bibnamefont {Siemensmeyer}},\ and\ \bibinfo {author} {\bibfnamefont
			{B.}~\bibnamefont {Lake}},\ }\bibfield  {title} {\bibinfo {title} {{Magnetic
				structure and crystal-field states of the pyrochlore antiferromagnet
				${\mathrm{Nd}}_{2}{\mathrm{Zr}}_{2}{\mathrm{O}}_{7}$}},\ }\href
	{https://doi.org/10.1103/PhysRevB.92.224430} {\bibfield  {journal} {\bibinfo
			{journal} {Phys. Rev. B}\ }\textbf {\bibinfo {volume} {92}},\ \bibinfo
		{pages} {224430} (\bibinfo {year} {2015})}\BibitemShut {NoStop}%
	\bibitem [{\citenamefont {Zhao}\ \emph {et~al.}(2022)\citenamefont {Zhao},
		\citenamefont {Ge}, \citenamefont {Zhou}, \citenamefont {Song}, \citenamefont
		{Yang}, \citenamefont {Li}, \citenamefont {Wang}, \citenamefont {Fu},
		\citenamefont {Zhang}, \citenamefont {Xu}, \citenamefont {Wang},
		\citenamefont {Mei}, \citenamefont {Tong}, \citenamefont {Wu},\ and\
		\citenamefont {Sheng}}]{Zhao014441}%
	\BibitemOpen
	\bibfield  {author} {\bibinfo {author} {\bibfnamefont {N.}~\bibnamefont
			{Zhao}}, \bibinfo {author} {\bibfnamefont {H.}~\bibnamefont {Ge}}, \bibinfo
		{author} {\bibfnamefont {L.}~\bibnamefont {Zhou}}, \bibinfo {author}
		{\bibfnamefont {Z.~M.}\ \bibnamefont {Song}}, \bibinfo {author}
		{\bibfnamefont {J.}~\bibnamefont {Yang}}, \bibinfo {author} {\bibfnamefont
			{T.~T.}\ \bibnamefont {Li}}, \bibinfo {author} {\bibfnamefont
			{L.}~\bibnamefont {Wang}}, \bibinfo {author} {\bibfnamefont {Y.}~\bibnamefont
			{Fu}}, \bibinfo {author} {\bibfnamefont {Y.~F.}\ \bibnamefont {Zhang}},
		\bibinfo {author} {\bibfnamefont {J.~B.}\ \bibnamefont {Xu}}, \bibinfo
		{author} {\bibfnamefont {S.~M.}\ \bibnamefont {Wang}}, \bibinfo {author}
		{\bibfnamefont {J.~W.}\ \bibnamefont {Mei}}, \bibinfo {author} {\bibfnamefont
			{X.}~\bibnamefont {Tong}}, \bibinfo {author} {\bibfnamefont {L.~S.}\
			\bibnamefont {Wu}},\ and\ \bibinfo {author} {\bibfnamefont {J.~M.}\
			\bibnamefont {Sheng}},\ }\bibfield  {title} {\bibinfo {title}
		{{Antiferromagnetism and Ising ground states in the rare-earth garnet
				${\mathrm{Nd}}_{3}{\mathrm{Ga}}_{5}{\mathrm{O}}_{12}$}},\ }\href
	{https://doi.org/10.1103/PhysRevB.105.014441} {\bibfield  {journal} {\bibinfo
			{journal} {Phys. Rev. B}\ }\textbf {\bibinfo {volume} {105}},\ \bibinfo
		{pages} {014441} (\bibinfo {year} {2022})}\BibitemShut {NoStop}%
	\bibitem [{\citenamefont {Cao}\ \emph {et~al.}(2025)\citenamefont {Cao},
		\citenamefont {Bu}, \citenamefont {Shiroka}, \citenamefont {Walker},
		\citenamefont {Fu}, \citenamefont {Tian}, \citenamefont {Zhao},\ and\
		\citenamefont {Guo}}]{Cao144409}%
	\BibitemOpen
	\bibfield  {author} {\bibinfo {author} {\bibfnamefont {Y.}~\bibnamefont
			{Cao}}, \bibinfo {author} {\bibfnamefont {H.}~\bibnamefont {Bu}}, \bibinfo
		{author} {\bibfnamefont {T.}~\bibnamefont {Shiroka}}, \bibinfo {author}
		{\bibfnamefont {H.~C.}\ \bibnamefont {Walker}}, \bibinfo {author}
		{\bibfnamefont {Z.}~\bibnamefont {Fu}}, \bibinfo {author} {\bibfnamefont
			{Z.}~\bibnamefont {Tian}}, \bibinfo {author} {\bibfnamefont {J.}~\bibnamefont
			{Zhao}},\ and\ \bibinfo {author} {\bibfnamefont {H.}~\bibnamefont {Guo}},\
	}\bibfield  {title} {\bibinfo {title} {{Magnetic ground state and persistent
				spin fluctuations in the triangular-lattice antiferromagnet
				${\mathrm{NdZnAl}}_{11}{\mathrm{O}}_{19}$}},\ }\href
	{https://doi.org/10.1103/tnwb-9hv8} {\bibfield  {journal} {\bibinfo
			{journal} {Phys. Rev. B}\ }\textbf {\bibinfo {volume} {112}},\ \bibinfo
		{pages} {144409} (\bibinfo {year} {2025})}\BibitemShut {NoStop}%
	\bibitem [{\citenamefont {Liu}\ \emph {et~al.}(2024)\citenamefont {Liu},
		\citenamefont {Song}, \citenamefont {Cao}, \citenamefont {Ge}, \citenamefont
		{Bu}, \citenamefont {Zhou}, \citenamefont {Qin}, \citenamefont {Zeng},
		\citenamefont {Li}, \citenamefont {Ling}, \citenamefont {Tong}, \citenamefont
		{Sheng}, \citenamefont {Yang}, \citenamefont {Wu}, \citenamefont {Guo},\ and\
		\citenamefont {Tian}}]{Liu184413}%
	\BibitemOpen
	\bibfield  {author} {\bibinfo {author} {\bibfnamefont {A.}~\bibnamefont
			{Liu}}, \bibinfo {author} {\bibfnamefont {F.}~\bibnamefont {Song}}, \bibinfo
		{author} {\bibfnamefont {Y.}~\bibnamefont {Cao}}, \bibinfo {author}
		{\bibfnamefont {H.}~\bibnamefont {Ge}}, \bibinfo {author} {\bibfnamefont
			{H.}~\bibnamefont {Bu}}, \bibinfo {author} {\bibfnamefont {J.}~\bibnamefont
			{Zhou}}, \bibinfo {author} {\bibfnamefont {Y.}~\bibnamefont {Qin}}, \bibinfo
		{author} {\bibfnamefont {Q.}~\bibnamefont {Zeng}}, \bibinfo {author}
		{\bibfnamefont {J.}~\bibnamefont {Li}}, \bibinfo {author} {\bibfnamefont
			{L.}~\bibnamefont {Ling}}, \bibinfo {author} {\bibfnamefont {W.}~\bibnamefont
			{Tong}}, \bibinfo {author} {\bibfnamefont {J.}~\bibnamefont {Sheng}},
		\bibinfo {author} {\bibfnamefont {M.}~\bibnamefont {Yang}}, \bibinfo {author}
		{\bibfnamefont {L.}~\bibnamefont {Wu}}, \bibinfo {author} {\bibfnamefont
			{H.}~\bibnamefont {Guo}},\ and\ \bibinfo {author} {\bibfnamefont
			{Z.}~\bibnamefont {Tian}},\ }\bibfield  {title} {\bibinfo {title} {{Distinct
				magnetic ground states in Shastry-Sutherland lattice materials:
				$\mathrm{P}{\mathrm{r}}_{2}\mathrm{B}{\mathrm{e}}_{2}\mathrm{Ge}{\mathrm{O}}_{7}$
				versus
				$\mathrm{N}{\mathrm{d}}_{2}\mathrm{B}{\mathrm{e}}_{2}\mathrm{Ge}{\mathrm{O}}_{7}$}},\
	}\href {https://doi.org/10.1103/PhysRevB.109.184413} {\bibfield  {journal}
		{\bibinfo  {journal} {Phys. Rev. B}\ }\textbf {\bibinfo {volume} {109}},\
		\bibinfo {pages} {184413} (\bibinfo {year} {2024})}\BibitemShut {NoStop}%
	\bibitem [{\citenamefont {Cussen}\ and\ \citenamefont
		{Yip}(2007)}]{Cussen1832}%
	\BibitemOpen
	\bibfield  {author} {\bibinfo {author} {\bibfnamefont {E.~J.}\ \bibnamefont
			{Cussen}}\ and\ \bibinfo {author} {\bibfnamefont {T.~W.}\ \bibnamefont
			{Yip}},\ }\bibfield  {title} {\bibinfo {title} {{A neutron diffraction study
				of the d0 and d10 lithium garnets Li$_3$Nd$_3$W$_2$O$_{12}$ and
				Li$_5$La$_3$Sb$_2$O$_{12}$}},\ }\href
	{https://doi.org/https://doi.org/10.1016/j.jssc.2007.04.007} {\bibfield
		{journal} {\bibinfo  {journal} {J. Solid State Chem.}\ }\textbf {\bibinfo
			{volume} {180}},\ \bibinfo {pages} {1832} (\bibinfo {year}
		{2007})}\BibitemShut {NoStop}%
	\bibitem [{\citenamefont {Rodríguez-Carvajal}(1993)}]{Carvajal55}%
	\BibitemOpen
	\bibfield  {author} {\bibinfo {author} {\bibfnamefont {J.}~\bibnamefont
			{Rodríguez-Carvajal}},\ }\bibfield  {title} {\bibinfo {title} {{Recent
				advances in magnetic structure determination by neutron powder
				diffraction}},\ }\href
	{https://doi.org/https://doi.org/10.1016/0921-4526(93)90108-I} {\bibfield
		{journal} {\bibinfo  {journal} {Physica B: Condensed Matter}\ }\textbf
		{\bibinfo {volume} {192}},\ \bibinfo {pages} {55} (\bibinfo {year}
		{1993})}\BibitemShut {NoStop}%
	\bibitem [{\citenamefont {Le}\ \emph {et~al.}(2023)\citenamefont {Le},
		\citenamefont {Guidi}, \citenamefont {Bewley}, \citenamefont {Stewart},
		\citenamefont {Schooneveld}, \citenamefont {Raspino}, \citenamefont {Pooley},
		\citenamefont {Boxall}, \citenamefont {Gascoyne}, \citenamefont {Rhodes},
		\citenamefont {Moorby}, \citenamefont {Templeman}, \citenamefont {Afford},
		\citenamefont {Waller}, \citenamefont {Zacek},\ and\ \citenamefont
		{Shaw}}]{Le168646}%
	\BibitemOpen
	\bibfield  {author} {\bibinfo {author} {\bibfnamefont {M.}~\bibnamefont
			{Le}}, \bibinfo {author} {\bibfnamefont {T.}~\bibnamefont {Guidi}}, \bibinfo
		{author} {\bibfnamefont {R.}~\bibnamefont {Bewley}}, \bibinfo {author}
		{\bibfnamefont {J.}~\bibnamefont {Stewart}}, \bibinfo {author} {\bibfnamefont
			{E.}~\bibnamefont {Schooneveld}}, \bibinfo {author} {\bibfnamefont
			{D.}~\bibnamefont {Raspino}}, \bibinfo {author} {\bibfnamefont
			{D.}~\bibnamefont {Pooley}}, \bibinfo {author} {\bibfnamefont
			{J.}~\bibnamefont {Boxall}}, \bibinfo {author} {\bibfnamefont
			{K.}~\bibnamefont {Gascoyne}}, \bibinfo {author} {\bibfnamefont
			{N.}~\bibnamefont {Rhodes}}, \bibinfo {author} {\bibfnamefont
			{S.}~\bibnamefont {Moorby}}, \bibinfo {author} {\bibfnamefont
			{D.}~\bibnamefont {Templeman}}, \bibinfo {author} {\bibfnamefont
			{L.}~\bibnamefont {Afford}}, \bibinfo {author} {\bibfnamefont
			{S.}~\bibnamefont {Waller}}, \bibinfo {author} {\bibfnamefont
			{D.}~\bibnamefont {Zacek}},\ and\ \bibinfo {author} {\bibfnamefont
			{R.}~\bibnamefont {Shaw}},\ }\bibfield  {title} {\bibinfo {title} {Upgrade of
			the mari spectrometer at isis},\ }\href
	{https://doi.org/https://doi.org/10.1016/j.nima.2023.168646} {\bibfield
		{journal} {\bibinfo  {journal} {Nuclear Instruments and Methods in Physics
				Research Section A: Accelerators, Spectrometers, Detectors and Associated
				Equipment}\ }\textbf {\bibinfo {volume} {1056}},\ \bibinfo {pages} {168646}
		(\bibinfo {year} {2023})}\BibitemShut {NoStop}%
	\bibitem [{\citenamefont {Arnold}\ \emph {et~al.}(2014)\citenamefont {Arnold},
		\citenamefont {Bilheux}, \citenamefont {Borreguero}, \citenamefont {Buts},
		\citenamefont {Campbell}, \citenamefont {Chapon}, \citenamefont {Doucet},
		\citenamefont {Draper}, \citenamefont {{Ferraz Leal}}, \citenamefont {Gigg},
		\citenamefont {Lynch}, \citenamefont {Markvardsen}, \citenamefont
		{Mikkelson}, \citenamefont {Mikkelson}, \citenamefont {Miller}, \citenamefont
		{Palmen}, \citenamefont {Parker}, \citenamefont {Passos}, \citenamefont
		{Perring}, \citenamefont {Peterson}, \citenamefont {Ren}, \citenamefont
		{Reuter}, \citenamefont {Savici}, \citenamefont {Taylor}, \citenamefont
		{Taylor}, \citenamefont {Tolchenov}, \citenamefont {Zhou},\ and\
		\citenamefont {Zikovsky}}]{Arnold156}%
	\BibitemOpen
	\bibfield  {author} {\bibinfo {author} {\bibfnamefont {O.}~\bibnamefont
			{Arnold}}, \bibinfo {author} {\bibfnamefont {J.}~\bibnamefont {Bilheux}},
		\bibinfo {author} {\bibfnamefont {J.}~\bibnamefont {Borreguero}}, \bibinfo
		{author} {\bibfnamefont {A.}~\bibnamefont {Buts}}, \bibinfo {author}
		{\bibfnamefont {S.}~\bibnamefont {Campbell}}, \bibinfo {author}
		{\bibfnamefont {L.}~\bibnamefont {Chapon}}, \bibinfo {author} {\bibfnamefont
			{M.}~\bibnamefont {Doucet}}, \bibinfo {author} {\bibfnamefont
			{N.}~\bibnamefont {Draper}}, \bibinfo {author} {\bibfnamefont
			{R.}~\bibnamefont {{Ferraz Leal}}}, \bibinfo {author} {\bibfnamefont
			{M.}~\bibnamefont {Gigg}}, \bibinfo {author} {\bibfnamefont {V.}~\bibnamefont
			{Lynch}}, \bibinfo {author} {\bibfnamefont {A.}~\bibnamefont {Markvardsen}},
		\bibinfo {author} {\bibfnamefont {D.}~\bibnamefont {Mikkelson}}, \bibinfo
		{author} {\bibfnamefont {R.}~\bibnamefont {Mikkelson}}, \bibinfo {author}
		{\bibfnamefont {R.}~\bibnamefont {Miller}}, \bibinfo {author} {\bibfnamefont
			{K.}~\bibnamefont {Palmen}}, \bibinfo {author} {\bibfnamefont
			{P.}~\bibnamefont {Parker}}, \bibinfo {author} {\bibfnamefont
			{G.}~\bibnamefont {Passos}}, \bibinfo {author} {\bibfnamefont
			{T.}~\bibnamefont {Perring}}, \bibinfo {author} {\bibfnamefont
			{P.}~\bibnamefont {Peterson}}, \bibinfo {author} {\bibfnamefont
			{S.}~\bibnamefont {Ren}}, \bibinfo {author} {\bibfnamefont {M.}~\bibnamefont
			{Reuter}}, \bibinfo {author} {\bibfnamefont {A.}~\bibnamefont {Savici}},
		\bibinfo {author} {\bibfnamefont {J.}~\bibnamefont {Taylor}}, \bibinfo
		{author} {\bibfnamefont {R.}~\bibnamefont {Taylor}}, \bibinfo {author}
		{\bibfnamefont {R.}~\bibnamefont {Tolchenov}}, \bibinfo {author}
		{\bibfnamefont {W.}~\bibnamefont {Zhou}},\ and\ \bibinfo {author}
		{\bibfnamefont {J.}~\bibnamefont {Zikovsky}},\ }\bibfield  {title} {\bibinfo
		{title} {{Mantid Data analysis and visualization package for neutron
				scattering and $\mu$SR experiments}},\ }\href
	{https://doi.org/https://doi.org/10.1016/j.nima.2014.07.029} {\bibfield
		{journal} {\bibinfo  {journal} {Nucl. Instrum. Methods Phys. Res. Sect. A}\
		}\textbf {\bibinfo {volume} {764}},\ \bibinfo {pages} {156} (\bibinfo {year}
		{2014})}\BibitemShut {NoStop}%
	\bibitem [{\citenamefont {Guchhait}\ \emph {et~al.}(2025)\citenamefont
		{Guchhait}, \citenamefont {Kolay}, \citenamefont {Magar},\ and\ \citenamefont
		{Nath}}]{Guchhait214437}%
	\BibitemOpen
	\bibfield  {author} {\bibinfo {author} {\bibfnamefont {S.}~\bibnamefont
			{Guchhait}}, \bibinfo {author} {\bibfnamefont {R.}~\bibnamefont {Kolay}},
		\bibinfo {author} {\bibfnamefont {A.}~\bibnamefont {Magar}},\ and\ \bibinfo
		{author} {\bibfnamefont {R.}~\bibnamefont {Nath}},\ }\bibfield  {title}
	{\bibinfo {title} {{{Magnetic and crystal electric field studies of the
					Yb$^{3+}$-based triangular lattice antiferromagnets NaSrYb(BO$_3$)$_2$ and
					K$_3$YbSi$_2$O$_7$}}},\ }\href {https://doi.org/10.1103/ks6z-6nxj} {\bibfield
		{journal} {\bibinfo  {journal} {Phys. Rev. B}\ }\textbf {\bibinfo {volume}
			{111}},\ \bibinfo {pages} {214437} (\bibinfo {year} {2025})}\BibitemShut
	{NoStop}%
	\bibitem [{\citenamefont {Sebastian}\ \emph {et~al.}(2025)\citenamefont
		{Sebastian}, \citenamefont {Kolay}, \citenamefont {B}, \citenamefont {Ding},
		\citenamefont {Furukawa},\ and\ \citenamefont {Nath}}]{Sebastian104428}%
	\BibitemOpen
	\bibfield  {author} {\bibinfo {author} {\bibfnamefont {S.~J.}\ \bibnamefont
			{Sebastian}}, \bibinfo {author} {\bibfnamefont {R.}~\bibnamefont {Kolay}},
		\bibinfo {author} {\bibfnamefont {A.}~\bibnamefont {B}}, \bibinfo {author}
		{\bibfnamefont {Q.-P.}\ \bibnamefont {Ding}}, \bibinfo {author}
		{\bibfnamefont {Y.}~\bibnamefont {Furukawa}},\ and\ \bibinfo {author}
		{\bibfnamefont {R.}~\bibnamefont {Nath}},\ }\bibfield  {title} {\bibinfo
		{title} {{Spin fluctuations, absence of magnetic order, and crystal electric
				field studies in the ${\mathrm{Yb}}^{3+}$-based triangular lattice
				antiferromagnet ${\mathrm{Rb}}_{3}\mathrm{Yb}{({\mathrm{VO}}_{4})}_{2}$}},\
	}\href {https://doi.org/10.1103/ydpn-8wgf} {\bibfield  {journal} {\bibinfo
			{journal} {Phys. Rev. B}\ }\textbf {\bibinfo {volume} {112}},\ \bibinfo
		{pages} {104428} (\bibinfo {year} {2025})}\BibitemShut {NoStop}%
	\bibitem [{\citenamefont {Kolay}\ \emph {et~al.}(2025)\citenamefont {Kolay},
		\citenamefont {Magar}, \citenamefont {Tsirlin},\ and\ \citenamefont
		{Nath}}]{Kolay104403}%
	\BibitemOpen
	\bibfield  {author} {\bibinfo {author} {\bibfnamefont {R.}~\bibnamefont
			{Kolay}}, \bibinfo {author} {\bibfnamefont {A.}~\bibnamefont {Magar}},
		\bibinfo {author} {\bibfnamefont {A.~A.}\ \bibnamefont {Tsirlin}},\ and\
		\bibinfo {author} {\bibfnamefont {R.}~\bibnamefont {Nath}},\ }\bibfield
	{title} {\bibinfo {title} {{Cluster-glass behavior and large magnetocaloric
				effect in the frustrated hyperkagome ferromagnet
				${\mathrm{Li}}_{2}{\mathrm{MgMn}}_{3}{\mathrm{O}}_{8}$}},\ }\href
	{https://doi.org/10.1103/PhysRevB.111.104403} {\bibfield  {journal} {\bibinfo
			{journal} {Phys. Rev. B}\ }\textbf {\bibinfo {volume} {111}},\ \bibinfo
		{pages} {104403} (\bibinfo {year} {2025})}\BibitemShut {NoStop}%
	\bibitem [{\citenamefont {Mugiraneza}\ and\ \citenamefont
		{Hallas}(2022)}]{Mugiraneza95}%
	\BibitemOpen
	\bibfield  {author} {\bibinfo {author} {\bibfnamefont {S.}~\bibnamefont
			{Mugiraneza}}\ and\ \bibinfo {author} {\bibfnamefont {A.~M.}\ \bibnamefont
			{Hallas}},\ }\bibfield  {title} {\bibinfo {title} {{Tutorial: a beginner's
				guide to interpreting magnetic susceptibility data with the Curie-Weiss
				law}},\ }\href {https://doi.org/10.1038/s42005-022-00853-y} {\bibfield
		{journal} {\bibinfo  {journal} {Commun. Phys.}\ }\textbf {\bibinfo {volume}
			{5}},\ \bibinfo {pages} {95} (\bibinfo {year} {2022})}\BibitemShut {NoStop}%
	\bibitem [{\citenamefont {Sebastian}\ \emph {et~al.}(2024)\citenamefont
		{Sebastian}, \citenamefont {Mohanty}, \citenamefont {Nath}, \citenamefont
		{Saravanan}, \citenamefont {Mandal}, \citenamefont {Tsirlin},\ and\
		\citenamefont {Nath}}]{Sebastian034403}%
	\BibitemOpen
	\bibfield  {author} {\bibinfo {author} {\bibfnamefont {S.~J.}\ \bibnamefont
			{Sebastian}}, \bibinfo {author} {\bibfnamefont {S.}~\bibnamefont {Mohanty}},
		\bibinfo {author} {\bibfnamefont {A.}~\bibnamefont {Nath}}, \bibinfo {author}
		{\bibfnamefont {M.~P.}\ \bibnamefont {Saravanan}}, \bibinfo {author}
		{\bibfnamefont {S.}~\bibnamefont {Mandal}}, \bibinfo {author} {\bibfnamefont
			{A.~A.}\ \bibnamefont {Tsirlin}},\ and\ \bibinfo {author} {\bibfnamefont
			{R.}~\bibnamefont {Nath}},\ }\bibfield  {title} {\bibinfo {title}
		{{Disordered ground state in a spin-orbit coupled pseudospin-$\frac{1}{2}$
				cobalt-based metal-organic framework magnet with orthogonal spin dimers}},\
	}\href {https://doi.org/10.1103/PhysRevMaterials.8.034403} {\bibfield
		{journal} {\bibinfo  {journal} {Phys. Rev. Mater.}\ }\textbf {\bibinfo
			{volume} {8}},\ \bibinfo {pages} {034403} (\bibinfo {year}
		{2024})}\BibitemShut {NoStop}%
	\bibitem [{\citenamefont {Kittel}(2004)}]{Kittel2004}%
	\BibitemOpen
	\bibfield  {author} {\bibinfo {author} {\bibfnamefont {C.}~\bibnamefont
			{Kittel}},\ }\href@noop {} {\emph {\bibinfo {title} {{Introduction to Solid
					State Physics}}}}\ (\bibinfo  {publisher} {Wiley, Hoboken, NJ},\ \bibinfo
	{year} {2004})\BibitemShut {NoStop}%
	\bibitem [{\citenamefont {Mohanty}\ \emph {et~al.}(2023)\citenamefont
		{Mohanty}, \citenamefont {Islam}, \citenamefont {Winterhalter-Stocker},
		\citenamefont {Jesche}, \citenamefont {Simutis}, \citenamefont {Wang},
		\citenamefont {Guguchia}, \citenamefont {Sichelschmidt}, \citenamefont
		{Baenitz}, \citenamefont {Tsirlin}, \citenamefont {Gegenwart},\ and\
		\citenamefont {Nath}}]{Mohanty134408}%
	\BibitemOpen
	\bibfield  {author} {\bibinfo {author} {\bibfnamefont {S.}~\bibnamefont
			{Mohanty}}, \bibinfo {author} {\bibfnamefont {S.~S.}\ \bibnamefont {Islam}},
		\bibinfo {author} {\bibfnamefont {N.}~\bibnamefont {Winterhalter-Stocker}},
		\bibinfo {author} {\bibfnamefont {A.}~\bibnamefont {Jesche}}, \bibinfo
		{author} {\bibfnamefont {G.}~\bibnamefont {Simutis}}, \bibinfo {author}
		{\bibfnamefont {C.}~\bibnamefont {Wang}}, \bibinfo {author} {\bibfnamefont
			{Z.}~\bibnamefont {Guguchia}}, \bibinfo {author} {\bibfnamefont
			{J.}~\bibnamefont {Sichelschmidt}}, \bibinfo {author} {\bibfnamefont
			{M.}~\bibnamefont {Baenitz}}, \bibinfo {author} {\bibfnamefont {A.~A.}\
			\bibnamefont {Tsirlin}}, \bibinfo {author} {\bibfnamefont {P.}~\bibnamefont
			{Gegenwart}},\ and\ \bibinfo {author} {\bibfnamefont {R.}~\bibnamefont
			{Nath}},\ }\bibfield  {title} {\bibinfo {title} {{Disordered ground state in
				the spin-orbit coupled ${J}_{\mathrm{eff}}$ = $\frac{1}{2}$ distorted
				honeycomb magnet ${\mathrm{BiYbGeO}}_{5}$}},\ }\href
	{https://doi.org/10.1103/PhysRevB.108.134408} {\bibfield  {journal} {\bibinfo
			{journal} {Phys. Rev. B}\ }\textbf {\bibinfo {volume} {108}},\ \bibinfo
		{pages} {134408} (\bibinfo {year} {2023})}\BibitemShut {NoStop}%
	\bibitem [{\citenamefont {Ahmed}\ \emph {et~al.}(2015)\citenamefont {Ahmed},
		\citenamefont {Tsirlin},\ and\ \citenamefont {Nath}}]{Ahmed214413}%
	\BibitemOpen
	\bibfield  {author} {\bibinfo {author} {\bibfnamefont {N.}~\bibnamefont
			{Ahmed}}, \bibinfo {author} {\bibfnamefont {A.~A.}\ \bibnamefont {Tsirlin}},\
		and\ \bibinfo {author} {\bibfnamefont {R.}~\bibnamefont {Nath}},\ }\bibfield
	{title} {\bibinfo {title} {{Multiple magnetic transitions in the
				spin-$\frac{1}{2}$ chain antiferromagnet
				${\mathrm{SrCuTe}}_{2}{\mathrm{O}}_{6}$}},\ }\href
	{https://doi.org/10.1103/PhysRevB.91.214413} {\bibfield  {journal} {\bibinfo
			{journal} {Phys. Rev. B}\ }\textbf {\bibinfo {volume} {91}},\ \bibinfo
		{pages} {214413} (\bibinfo {year} {2015})}\BibitemShut {NoStop}%
	\bibitem [{\citenamefont {Nath}\ \emph {et~al.}(2008)\citenamefont {Nath},
		\citenamefont {Tsirlin}, \citenamefont {Rosner},\ and\ \citenamefont
		{Geibel}}]{Nath064422}%
	\BibitemOpen
	\bibfield  {author} {\bibinfo {author} {\bibfnamefont {R.}~\bibnamefont
			{Nath}}, \bibinfo {author} {\bibfnamefont {A.~A.}\ \bibnamefont {Tsirlin}},
		\bibinfo {author} {\bibfnamefont {H.}~\bibnamefont {Rosner}},\ and\ \bibinfo
		{author} {\bibfnamefont {C.}~\bibnamefont {Geibel}},\ }\bibfield  {title}
	{\bibinfo {title} {{Magnetic properties of
				$\text{BaCdVO}{({\text{PO}}_{4})}_{2}$: A strongly frustrated
				spin-$\frac{1}{2}$ square lattice close to the quantum critical regime}},\
	}\href {https://doi.org/10.1103/PhysRevB.78.064422} {\bibfield  {journal}
		{\bibinfo  {journal} {Phys. Rev. B}\ }\textbf {\bibinfo {volume} {78}},\
		\bibinfo {pages} {064422} (\bibinfo {year} {2008})}\BibitemShut {NoStop}%
	\bibitem [{INS()}]{INS_Data_Ref}%
	\BibitemOpen
	\href@noop {} {}\bibinfo {howpublished}
	{\url{https://doi.org/10.5286/ISIS.E.RB2610598-1}}\BibitemShut {NoStop}%
	\bibitem [{\citenamefont {Boothroyd}(2020)}]{Boothroyd2020}%
	\BibitemOpen
	\bibfield  {author} {\bibinfo {author} {\bibfnamefont {A.}~\bibnamefont
			{Boothroyd}},\ }\href {https://books.google.dk/books?id=FvTuDwAAQBAJ} {\emph
		{\bibinfo {title} {Principles of Neutron Scattering from Condensed Matter}}}\
	(\bibinfo  {publisher} {OUP Oxford},\ \bibinfo {year} {2020})\BibitemShut
	{NoStop}%
	\bibitem [{\citenamefont {Stevens}(1952)}]{Stevens209}%
	\BibitemOpen
	\bibfield  {author} {\bibinfo {author} {\bibfnamefont {K.~W.~H.}\
			\bibnamefont {Stevens}},\ }\bibfield  {title} {\bibinfo {title} {{Matrix
				Elements and Operator Equivalents Connected with the Magnetic Properties of
				Rare Earth Ions}},\ }\href {https://doi.org/10.1088/0370-1298/65/3/308}
	{\bibfield  {journal} {\bibinfo  {journal} {Proc. Phys. Soc. Section A}\
		}\textbf {\bibinfo {volume} {65}},\ \bibinfo {pages} {209} (\bibinfo {year}
		{1952})}\BibitemShut {NoStop}%
	\bibitem [{\citenamefont {Hutchings}(1964)}]{Huthings227}%
	\BibitemOpen
	\bibfield  {author} {\bibinfo {author} {\bibfnamefont {M.}~\bibnamefont
			{Hutchings}},\ }\href
	{https://doi.org/https://doi.org/10.1016/S0081-1947(08)60517-2} {\emph
		{\bibinfo {title} {{Point-Charge Calculations of Energy Levels of Magnetic
					Ions in Crystalline Electric Fields}}}},\ \bibinfo {series} {Solid State
		Physics}, Vol.~\bibinfo {volume} {16}\ (\bibinfo  {publisher} {Academic
		Press},\ \bibinfo {year} {1964})\ p.\ \bibinfo {pages} {227}\BibitemShut
	{NoStop}%
	\bibitem [{\citenamefont {Newman}\ and\ \citenamefont {Ng}(2000)}]{Newman2000}%
	\BibitemOpen
	\bibfield  {author} {\bibinfo {author} {\bibfnamefont {D.~J.}\ \bibnamefont
			{Newman}}\ and\ \bibinfo {author} {\bibfnamefont {B.}~\bibnamefont {Ng}},\
	}\href {https://doi.org/https://doi.org/10.1017/CBO9780511524295} {\emph
		{\bibinfo {title} {Crystal Field Handbook}}}\ (\bibinfo  {publisher}
	{Cambridge University Press},\ \bibinfo {year} {2000})\BibitemShut {NoStop}%
	\bibitem [{\citenamefont {Kutuzov}\ and\ \citenamefont
		{Skvortsova}(2011)}]{Kutuzov012039}%
	\BibitemOpen
	\bibfield  {author} {\bibinfo {author} {\bibfnamefont {A.~S.}\ \bibnamefont
			{Kutuzov}}\ and\ \bibinfo {author} {\bibfnamefont {A.~M.}\ \bibnamefont
			{Skvortsova}},\ }\bibfield  {title} {\bibinfo {title} {{Crystal electric
				field parameters for Yb$^{3+}$ ion in YbRh$_2$Si$_2$}},\ }\href
	{https://doi.org/10.1088/1742-6596/324/1/012039} {\bibfield  {journal}
		{\bibinfo  {journal} {J. Phys.: Confer. Ser.}\ }\textbf {\bibinfo {volume}
			{324}},\ \bibinfo {pages} {012039} (\bibinfo {year} {2011})}\BibitemShut
	{NoStop}%
	\bibitem [{\citenamefont {Xiang}\ \emph {et~al.}()\citenamefont {Xiang},
		\citenamefont {Su}, \citenamefont {Xi}, \citenamefont {Fu}, \citenamefont
		{Chen}, \citenamefont {Jin}, \citenamefont {Chen}, \citenamefont {Mo},
		\citenamefont {Qi}, \citenamefont {Shen}, \citenamefont {Zhang},
		\citenamefont {Jin}, \citenamefont {Li}, \citenamefont {Sun},\ and\
		\citenamefont {Su}}]{Xiang2023}%
	\BibitemOpen
	\bibfield  {author} {\bibinfo {author} {\bibfnamefont {J.}~\bibnamefont
			{Xiang}}, \bibinfo {author} {\bibfnamefont {C.}~\bibnamefont {Su}}, \bibinfo
		{author} {\bibfnamefont {N.}~\bibnamefont {Xi}}, \bibinfo {author}
		{\bibfnamefont {Z.}~\bibnamefont {Fu}}, \bibinfo {author} {\bibfnamefont
			{Z.}~\bibnamefont {Chen}}, \bibinfo {author} {\bibfnamefont {H.}~\bibnamefont
			{Jin}}, \bibinfo {author} {\bibfnamefont {Z.}~\bibnamefont {Chen}}, \bibinfo
		{author} {\bibfnamefont {Z.-J.}\ \bibnamefont {Mo}}, \bibinfo {author}
		{\bibfnamefont {Y.}~\bibnamefont {Qi}}, \bibinfo {author} {\bibfnamefont
			{J.}~\bibnamefont {Shen}}, \bibinfo {author} {\bibfnamefont {L.}~\bibnamefont
			{Zhang}}, \bibinfo {author} {\bibfnamefont {W.}~\bibnamefont {Jin}}, \bibinfo
		{author} {\bibfnamefont {W.}~\bibnamefont {Li}}, \bibinfo {author}
		{\bibfnamefont {P.}~\bibnamefont {Sun}},\ and\ \bibinfo {author}
		{\bibfnamefont {G.}~\bibnamefont {Su}},\ }\bibfield  {title} {\bibinfo
		{title} {{Dipolar Spin Liquid Ending with Quantum Critical Point in a
				Gd-based Triangular Magnet}},\ }\href {https://arxiv.org/abs/2301.03571} {\
	}\Eprint {https://arxiv.org/abs/2301.03571} {arXiv:2301.03571} \BibitemShut
	{NoStop}%
	\bibitem [{\citenamefont {Guo}\ \emph {et~al.}(2019)\citenamefont {Guo},
		\citenamefont {Ghasemi}, \citenamefont {Broholm},\ and\ \citenamefont
		{Cava}}]{Guo094404}%
	\BibitemOpen
	\bibfield  {author} {\bibinfo {author} {\bibfnamefont {S.}~\bibnamefont
			{Guo}}, \bibinfo {author} {\bibfnamefont {A.}~\bibnamefont {Ghasemi}},
		\bibinfo {author} {\bibfnamefont {C.~L.}\ \bibnamefont {Broholm}},\ and\
		\bibinfo {author} {\bibfnamefont {R.~J.}\ \bibnamefont {Cava}},\ }\bibfield
	{title} {\bibinfo {title} {Magnetism on ideal triangular lattices in
			$\mathrm{NaBaYb}{(\mathrm{B}{\mathrm{O}}_{3})}_{2}$},\ }\href
	{https://doi.org/10.1103/PhysRevMaterials.3.094404} {\bibfield  {journal}
		{\bibinfo  {journal} {Phys. Rev. Mater.}\ }\textbf {\bibinfo {volume} {3}},\
		\bibinfo {pages} {094404} (\bibinfo {year} {2019})}\BibitemShut {NoStop}%
\end{thebibliography}

%apsrev4-2.bst 2019-01-14 (MD) hand-edited version of apsrev4-1.bst
%Control: key (0)
%Control: author (8) initials jnrlst
%Control: editor formatted (1) identically to author
%Control: production of article title (0) allowed
%Control: page (0) single
%Control: year (1) truncated
%Control: production of eprint (0) enabled
%

\end{document}